\documentclass[twocolumn]{autart}  

\usepackage{amssymb,latexsym,amsfonts,amsmath}

\usepackage{mathrsfs}

\usepackage{natbib}

\usepackage{xspace}

\let\oldemptyset\emptyset
\usepackage{algorithm}
\usepackage{algorithmic}
\newcommand{\redsquare}{\tikz\fill[red!100!black] (0,0) rectangle (2mm,2mm);}
\newcommand{\bluesquare}{\tikz\fill[blue!40!white] (0,0) rectangle (2mm,2mm);}
\newcommand{\greensquare}{\tikz\fill[green!70!black] (0,0) rectangle (2mm,2mm);}
\newcommand{\pinksquare}{\tikz\fill[pink!80!white] (0,0) rectangle (2mm,2mm);}
\usepackage{multirow}
\usepackage{booktabs}
\usepackage[dvipsnames]{xcolor}

\usepackage{url}  

\usepackage[
colorlinks=true,
linkcolor=NavyBlue,
citecolor=NavyBlue,
urlcolor=Goldenrod
]{hyperref}

\makeatletter
\edef\endfrontmatter{%
	\unexpanded\expandafter{\endfrontmatter}
	\noexpand\endNoHyper                  
}
\makeatother

\DeclareRobustCommand\sampleline[1]{%
	\tikz\draw[#1] (0,0) (0,\the\dimexpr\fontdimen22\textfont2\relax)
	-- (2em,\the\dimexpr\fontdimen22\textfont2\relax);%
}

\usepackage{caption}

\usepackage{graphicx}
\usepackage{paralist}
\usepackage{dsfont}
\usepackage{caption} 
\usepackage{booktabs}
\usepackage{eso-pic}
\usepackage{epsfig}
\usepackage{mathtools}
\usepackage {url}
\usepackage{epstopdf}
\usepackage{tikz}
\usepackage{IEEEtrantools}
\usetikzlibrary{arrows,calc,decorations.markings,math,arrows.meta}
\newtheorem{theorem}{Theorem}[section]
\newtheorem{lemma}[theorem]{Lemma}
\newtheorem{problem}[theorem]{Problem}

\newtheorem{definition}[theorem]{Definition}

\newtheorem{remark}[theorem]{Remark}
\newtheorem{assumption}[theorem]{Assumption}
\numberwithin{equation}{section}

\newcommand{\R}{{\mathbb{R}}}

\usepackage{dsfont}
\usepackage{caption}

\usepackage[many]{tcolorbox}
\usetikzlibrary{calc}
\tcbuselibrary{skins}

\newtcolorbox{resp}[1][]{%
	enhanced jigsaw,%
	colback=gray!5!white,%
	colframe=gray!80!black,%
	size=small,%
	boxrule=1pt,%
	halign title=flush center,%
	coltitle=black,%
	breakable,%
	drop shadow=black!50!white,%
	attach boxed title to top left={xshift=1cm,yshift=-\tcboxedtitleheight/2,yshifttext=-\tcboxedtitleheight/2},%
	minipage boxed title=3cm,%
	boxed title style={%
		colback=white,%
		size=fbox,%
		boxrule=1pt,%
		boxsep=2pt,%
		underlay={%
			\coordinate (dotA) at ($(interior.west) + (-0.5pt,0)$);
			\coordinate (dotB) at ($(interior.east) + (0.5pt,0)$);
			\begin{scope}[gray!80!black]
				\fill (dotA) circle (2pt);
				\fill (dotB) circle (2pt);
			\end{scope}
		}%
	},%
	#1%
}

\newcommand{\x}{\mathbf{x}}

\newcommand{\U}{{\mathbf{U}}}

\usepackage{subcaption}

\usepackage[bb=boondox]{mathalfa}

\usepackage{enumitem}

\makeatletter
\long\def\@maketablecaption#1#2{\@tablecaptionsize
    \global \@minipagefalse
    \hbox to \hsize{\parbox[t]{\hsize}{\centering #1 \\ #2}}}

\makeatother

\begin{document}

\begin{frontmatter}
\title{From Data to Control: A Formal Compositional Framework for Large-Scale Interconnected Networks}
\author{Omid Akbarzadeh}\ead{omid.akbarzadeh@newcastle.ac.uk},
\author{Amy Nejati}\ead{amy.nejati@newcastle.ac.uk}, and  
\author{Abolfazl Lavaei}\ead{abolfazl.lavaei@newcastle.ac.uk}  
\address{School of Computing, Newcastle University, United Kingdom}                           


\begin{abstract}  
We introduce a \emph{compositional data-driven} methodology with noisy data for designing \emph{fully-decentralized} safety controllers applicable to large-scale interconnected networks, encompassing a vast number of subsystems with \emph{unknown} mathematical models. Our compositional scheme leverages the interconnection topology and breaks down the network analysis into the examination of distinct subsystems. This is accompanied by utilizing a concept of \emph{control storage certificates} (CSCs) to capture joint dissipativity-type properties among subsystems. These CSCs are instrumental in a \emph{compositional} derivation of a \emph{control barrier certificate (CBC)} specialized for the interconnected network, thereby ensuring its safety. In our data-driven scheme, we gather only a single noise-corrupted \emph{input-state trajectory} from each unknown subsystem within a specified time frame. By fulfilling a specific rank condition, this process facilitates the construction of a CSC for each subsystem. Following this, by adhering to \emph{compositional dissipativity} reasoning, we compose CSCs derived from noisy data and build a CBC for the unknown network, ensuring its safety over an \emph{infinite time horizon}, while providing \emph{correctness guarantees}. We demonstrate that our compositional data-driven approach significantly enhances the design of a CBC and its robust safety controller under noisy data across the interconnected network. This advancement is achieved by reducing the computational complexity from a \emph{polynomial growth} in relation to network dimension, when using sum-of-squares (SOS) optimization, to a \emph{linear scale} based on the number of subsystems. We additionally demonstrate that the dissipativity-type compositionality condition can benefit from the structure of interconnection topology and potentially be fulfilled regardless of the number of subsystems.  We apply our data-driven findings to a variety of benchmarks, involving physical networks with unknown models and diverse interconnection topologies.
\end{abstract}

\end{frontmatter}

\section{Introduction}\label{sec:intro}\vspace{-0.2cm}
Recent advancements in data-driven methods have proven to be a formidable alternative for circumventing the conventional model-based techniques. These approaches employ system measurements to conduct safety analysis across diverse real-world applications, where precise mathematical models are either unavailable or too complex for practical implementation. Especially in scenarios where traditional model-based approaches struggle with analyzing complex systems, data-driven methods have emerged in two categories: \emph{indirect and direct}, both showing significant potential across various application domains~\citep{dorfler2022bridging,Hou2013model,BISOFFI20203953}. 

\emph{Indirect} data-driven methods typically involve system identification followed by model-based controller analysis. The primary advantage of this approach lies in its capability to utilize the extensive tools, developed as model-based control techniques, following the completion of the identification phase. Nonetheless, a significant limitation entails the computational complexity encountered in \emph{two phases}: initially in model identification, and subsequently in solving the model-based problem. On the other hand, \emph{direct} data-driven methods bypass the system identification step and directly employ system measurements to provide analyses across unknown systems (\emph{e.g.,} learning control policies directly from data~\citep{Indrict-data-driven, Indrict-data-driven1,samari2024single,samari2025data,zaker2025data} and stability analysis using data-driven methods in the presence of noisy  data~\citep{De-Persis-noisy,de2019formulas}).  

Transitioning to formal verification and controller synthesis of complex dynamical systems, this area has witnessed a surge in popularity, with efforts concentrated on fulfilling high-level logic specifications, often articulated through linear temporal logic (LTL) formulae~\citep{baier2008principles}. This growing interest is driven by the need to ensure safety and reliability requirements in dynamically complex systems, where alternative (traditional) approaches often fall short. In particular, the challenge primarily escalates when dealing with \emph{continuous-space} systems, a common occurrence in numerous real-world scenarios. To mitigate this difficulty, the relevant literature has predominantly relied on finite abstractions to simplify original models into \emph{discrete-space} representations~\citep{tabuada2009verification,julius2009approximations,zamani2016approximations,belta2017formal,lavaei2022scalable,nejati2020compositional,lavaei2024abstraction}; however, this approach faces significant computational complexity due to the \emph{exponential} state-explosion problem.

To address the computational burden posed by discretization-based methods, a promising alternative involves employing \emph{control barrier certificates (CBC)}, as a discretization-free approach, initially introduced by~\cite{Pranja}. 
In particular, barrier certificates resemble Lyapunov functions, serving to fulfill particular conditions concerning the function itself and its evolution within the system's dynamics. Through the establishment of a level set derived from the system's initial states, it effectively separates unsafe regions from the system's trajectories, consequently offering a formal (probabilistic) assurance across the system's safety~\citep{ames2019control,clark2021control,Mahathi, NEURIPS2020_barrier,lavaei2024scalable,zaker2024compositional,nejati2024context}.

While barrier certificates hold significant promise in the safety analysis of dynamical systems without requiring discretization, they still encounter two main challenges. Firstly, existing tools for searching for a CBC, such as SOS programming, are not scalable with the dimension of dynamical systems, rendering them computationally intractable for large-scale networks. Secondly, existing methods require precise knowledge of the system model when searching for a CBC, whereas in practical applications, the system dynamics are frequently unknown. While recent endeavors have aimed to design a CBC using data, current approaches are only suitable for low-dimensional systems and face significant challenges, with either polynomial~\citep{bisoffi2020controller,nejati2022data} or exponential~\citep{nejati2023data} sample complexity with respect to the state size when applied to large-scale systems.

{\bf Key contributions.} Inspired by these two primary challenges, our paper offers a data-driven approach with noisy data in a \emph{compositional framework} aimed at enforcing safety specifications across \emph{large-scale interconnected networks} with unknown dynamics. By leveraging joint dissipativity-type properties among subsystems and constructing \emph{control storage certificates} (CSCs) from data, we synthesize local controllers for each subsystem based solely on a \emph{single noise-corrupted input-state trajectory}, fulfilling a specific rank condition according to the persistency of excitation~\citep{persistency}. While adhering to some compositional conditions grounded in \emph{dissipativity reasoning}, we then construct a control barrier certificate for the interconnected network alongside its \emph{fully-decentralized} robust safety controller under noisy data via data-driven CSCs of individual subsystems, thereby guaranteeing network safety. As the network dimension increases, the computational complexity for constructing a CBC and its controller using SOS optimization escalates at a \emph{polynomial} rate. In contrast, our compositional data-driven method significantly reduces this complexity to a \emph{linear} level concerning the number of subsystems. We showcase the efficacy of our data-driven findings across a range of physical benchmarks with unknown models and diverse interconnection topologies.

\section{Problem formulation}{\label{sec:Continuous}}\vspace{-0.4cm}
\subsection{Notation}
Symbols $\mathbb{R}$, $\mathbb{R}^{+}_0$, and $\mathbb{R}^{+}$ represent sets of real numbers, non-negative and positive real numbers, respectively. Sets of non-negative and positive integers are, respectively, denoted by $\mathbb{N}:= \{0, 1, 2, \ldots\}$ and $\mathbb{N}^{+}:= \{1, 2, \ldots\}$. For $N$ vectors $x_i \in \mathbb{R}^{n_i}$, the notation $x = [x_1; \ldots; x_N]$ is employed to signify a column vector formed by these vectors, with a dimension of $\sum_i n_i$. Moreover, we use $[x_1\, \ldots\, x_N]$ to represent the horizontal concatenation of vectors $x_i \in \mathbb{R}^n$ to form $n \times N$ matrix. The notation $\{a_{ij}\}$ represents a matrix formed by placing the elements $a_{ij}$ in the $i$-th row and $j$-th column. The Euclidean norm of a vector $x \in \mathbb{R}^n$ is expressed as $\Vert x \Vert$. For sets $X_i, i \in \{1, \ldots, N\}$, their Cartesian product is denoted as $\prod_{i=1}^N X_i$. The identity matrix in $\mathbb{R}^{n \times n}$ is represented by $\mathbf{I}_n$. A zero matrix of dimension $n\times m$ is denoted by  $\mathbf{0}_{n\times m}$, while $\mathbf{0}_{n}$ represents the zero \emph{vector} of dimension $n$. A \emph{symmetric} and positive-definite matrix $P\in \mathbb{R}^{n \times n} $ is denoted by $P \succ 0$, while $P \succeq 0$ represents that $P$ is a \emph{symmetric} positive semi-definite matrix. A block diagonal matrix in $\R^{N\times{N}}$ with  diagonal matrix entries $(A_1,\ldots,A_N)$ is signified by $\mathsf{blkdiag}(A_1,\ldots,A_N)$. We denote the empty set as $\oldemptyset$. The transpose of a matrix $P$ is represented by $P^\top$\!, while its \emph{left} pseudoinverse is denoted by $P^{\dagger}$.\vspace{-0.2cm}

\subsection{Individual subsystems}\label{systems1}\vspace{-0.2cm}
We concentrate on continuous-time nonlinear polynomial systems, considering them as \emph{subsystems} in accordance with the forthcoming definition.
\begin{definition}
A continuous-time nonlinear polynomial system (ct-NPS) is characterized by
\begin{subequations} 
\begin{align}\label{sys2}
\Xi_i\!:
\dot x_i=A_i \mathcal R_i (x_i) + B_iu_i + D_iw_i,
\end{align}
where $A_i \in \mathbb R^{n_i\times {M_i}}, B_i \in \mathbb R^{n_i\times m_i},D_i \in \mathbb R^{n_i\times n_i},$ and  $\mathcal R_i(x_i) \in \mathbb R^{M_i}$, with $\mathcal{R}_i(\boldsymbol{0}_{n_i})=\boldsymbol{0}_{M_i}$, is a vector of monomials in states $x_i\in  \mathbb R^{n_i}$ fulfilling
\begin{equation}\label{Transformation}
	\mathcal{R}_i(x_i)= \Theta_i(x_i) x_i,
\end{equation}
 with $ \Theta_i(x_i) \in \mathbb{R}^{M_i \times n_i}$ being a state-dependent transformation matrix. Furthermore, $u_i \in U_i$ and $w_i \in W_i$ are external and internal inputs of ct-NPS, with $X_i \subseteq \mathbb R^{n_i}$, $U_i \subseteq \mathbb R^{m_i}$, and $W_i \subseteq \mathbb{R}^{n_i}$ being state, external, and internal input sets, respectively. We represent ct-NPS in \eqref{sys2} using the tuple $\Xi_i = \left(A_i, B_i, D_i, X_i, U_i, W_i\right)\!.$
 
A ct-NPS in~\eqref{sys2} can be equivalently  rewritten as
\begin{align}\label{sys3}
	\Xi_i\!: \dot x_i=A_i \Theta_i(x_i)x_i + B_i u_i + D_i w_i.
\end{align}
\end{subequations} 

\end{definition}
\begin{remark}
The transformation $\Theta_i(x_i)$ introduced in~\eqref{Transformation} ensures that all expressions depend directly on $x_i$, rather than $\mathcal{R}_i(x_i)$, simplifying the analysis and facilitating the development of relaxed conditions as proposed in Theorem~\ref{Thm:main2}. This approach is also consistent with our choice of the CSC defined later as $\mathcal{S}_i(x_i) = x_i^\top P_i x_i$, which explicitly depends only on $x_i$. It is worth noting that, without loss of generality, for any $\mathcal{R}_i(x_i)$ satisfying $\mathcal{R}_i(\boldsymbol{0}_{n_i})=\boldsymbol{0}_{M_i}$, there exists a transformation $\Theta_i(x_i)$ fulfilling~\eqref{Transformation}.
\end{remark}

In our setting, both matrices $A_i$ and $B_i$ are \emph{unknown}, while we assume knowledge of matrix $D_i$, as it represents the interconnection weights of subsystems within the topology, which is typically available in interconnected networks. We also assume access to an extensive form of $\mathcal{R}_i(x_i)$.

\begin{remark}
While the exact form of the dictionary $\mathcal{R}_i(x_i)$—referring to a library or family of functions— remains unknown, we assume access to a  sufficiently extensive monomials capable of capturing the actual system dynamics. In particular, this dictionary should be comprehensive enough to encompass all relevant terms that could appear in the actual dynamics, even if some of these terms turn out to be superfluous. In practical scenarios, one can consider an upper bound on the maximum degree of the monomial terms of $\mathcal{R}_i(x_i)$ based on the system's physical insights, ensuring that all state combinations up to that bound are included in $\mathcal{R}_i(x_i)$ (cf. case studies).
\end{remark}

\subsection{Interconnected networks}\label{In-Net}
To assess the safety of interconnected networks built from individual subsystems defined in \eqref{sys2}, this subsection details the network's configuration and describes how these subsystems are interconnected.
Here, we present the formal definition of interconnected networks composed of numerous individual ct-NPS.
\begin{definition}\label{network}
Consider $N \in \mathbb{N}^{+}$ subsystems $\Xi_i=\left(A_i, B_i, D_i, X_i, U_i, W_i\right)$, together with a coupling block matrix $\mathds{M}=\{\mathbb{m}_{ij}\}$, $i,j \in\{1, \ldots, N\}$, with binary entries, outlining the interaction among subsystems as the topology. The interconnection of subsystems $\Xi_i$, signified by $\mathcal{I}\left(\Xi_1, \ldots, \Xi_N\right)$, constitutes the interconnected network $\Xi=(A(x), B, X, U)$, where $X:=\prod_{i=1}^N X_i$, and $U:=\prod_{i=1}^N U_i$, with the internal-input constraint adhering to
\begin{equation}\label{topology}
   \left[w_1 ; \ldots ; w_N\right]=\mathds{M}\left[x_1 ; \ldots ; x_N\right]\!. 
\end{equation}
Such an interconnected network can be characterized by
\begin{align}\label{sysN}
		\Xi\!:\dot x=A(x)x + Bu,
\end{align}
where $A(x) =\left\{\mathbb{a}_{i j}\right\}$, $i,j \in\{1, \ldots, N\},$  is a block matrix with diagonal blocks $(A_1\Theta_1(x_1), \ldots, A_N\Theta_N(x_N))$, and off-diagonal blocks either derived from ${D}_{i}$ or set to zero, depending on the interconnection topology. Furthermore, $B = \mathsf{blkdiag}(B_1, \ldots, B_N) \in \mathbb{R}^{n \times m}$, with $n= \sum_{i=1}^N n_i$ and $m= \sum_{i=1}^N m_i$, while $u = [u_1; \dots; u_N] \in \mathbb{R}^{m}$ and $x = [x_1; \dots; x_N] \in \mathbb{R}^{n}$.
\end{definition}

We denote the value of the solution process of $\Xi$ at time $t \in \mathbb{R}^{+}_{0}$ with $x_{x_{0,u}}(t)$, under an input trajectory $u(\cdot)$ and starting from any initial state $x_0\in X$.  

\begin{remark}
In general, the interconnection topology $\mathds{M}$ may contain non-binary entries. In such cases, the off-diagonal blocks of $A(x)$ in~\eqref{sysN} can be expressed as $D_i \mathbb{m}_{ij}$ for $i \neq j$, where $\mathbb{m}_{ij}$ may take arbitrary non-binary values. Additionally, we assume that the diagonal blocks of \(\mathds{M}\) are $\mathbf{0}_{n_i \times n_i}$, implying that the subsystems have no self-connections. However, this assumption can be generalized by allowing subsystems to have self-loops. In this case, the diagonal blocks of \(A(x)\) are represented as \((A_1\Theta_1(x_1) + D_{1}, \ldots, A_N\Theta_N(x_N) + D_N)\).
\end{remark}

In this work, the term ``internal'' pertains to inputs of subsystems that influence each other within the interconnection topology, whereby the states from one subsystem affect an internal input of another one (cf. the interconnection constraint \eqref{topology}). Conversely, we employ ``external'' to denote inputs not engaged in forming the interconnection. The safety property is defined over the states of the interconnected network, and the primary objective is to synthesize external inputs that enforce such a property. While stability analysis of interconnected networks via dissipativity typically requires no bounds on internal inputs \citep{arcak2016networks}, in our work, these inputs are bounded since they are functions of the neighboring subsystem states (cf.~\eqref{topology}), and the state space is compact due to the focus on safety rather than stability.

\begin{remark}
Our framework functions under instantaneous interconnection, meaning that at any given moment, each subsystem can be influenced simultaneously by internal inputs. In addition, we assume lossless data exchange along the interconnection links and neglect communication delays. While our previous work investigated the effects of data loss in communication links within control systems \citep{Omid-wireless,Omid-wireless-data}, we do not consider it here for the sake of readability. Nevertheless, these studies offer a foundational basis for extending the current framework to handle data loss or time delays, as a promising direction for future research.
\end{remark}
    
\subsection{Control storage  and barrier certificates}\vspace{-0.2cm}
This subsection aims to ensure safety across an interconnected network by presenting the concept of \emph{control storage and barrier certificates}, for both individual subsystems and interconnected networks, with and without internal signals, respectively.
In the following, we first introduce the notion of control storage certificates for ct-NPS with \emph{internal} inputs.
\begin{definition}\label{barrier-condirions-model}
Consider a ct-NPS ~$\Xi_i=(A_i, B_i,D_i, X_i,\\U_i, W_i)$, $ i \in\{1, \ldots, N\}$, and sets $X_{0_i}, X_{a_i} \subseteq X_i$, where the initial set $X_{0_i}$ comprises states from which the subsystem $\Xi_i$ can initiate its operation, while the unsafe set $X_{a_i}$ encompasses states that must be avoided due to safety concerns. Assuming the existence of constants $\eta_i,\mu_i,\lambda_i\in\R^{+}$, and a symmetric block matrix $\mathds Z_i$, partitioned into $\mathds Z_i^{rr'}$, $r,r'\in\{1,2\}$,  with $ \mathds Z^{22}_i \preceq 0$, a function $\mathcal S_i:X_i\rightarrow\mathbb{R}_{0}^{+}$ is called a control storage certificate (CSC) for $\Xi_i$, if
\begin{subequations}
	\begin{align}\label{subsys1}
		&\mathcal S_i(x_i) \leq \eta_i,\quad\quad\quad\quad\quad\quad\!\forall x_i \in X_{0_i},\\\label{subsys2}
		&\mathcal S_i(x_i) \geq \mu_i, \quad\quad\quad\quad\quad\quad\!\forall x_i \in X_{a_i}, 
	\end{align}
and $\forall x_i\in X_i$, $\exists u_i\in U_i$, such that $\forall w_i\in W_i$, one has 
\begin{align}\label{eq:martingale2}
	&\mathsf{L}\mathcal S_i(x_i)\mathcal ~\leq - \lambda_i \mathcal S_i(x_i) \! +\! \begin{bmatrix}
		w_i\\
		x_i
	\end{bmatrix}^\top\!
	{\underbrace{\begin{bmatrix}
			\mathds Z^{11}_i& \mathds  Z^{12}_i\\
			\mathds Z^{21}_i& \mathds  Z^{22}_i
	\end{bmatrix}}_{\mathds Z_i}}\!\begin{bmatrix}
		w_i\\
	   x_i
	\end{bmatrix}\!,
\end{align}
\end{subequations}
 where $\lambda_i$ denotes the decay rate, $\mathds Z^{21}_i = \mathds Z_i^{{12}^\top}$, and $\mathsf{L} \mathcal S_i$ is the Lie derivative~\citep{Willmore_1960} of $\mathcal S_i$
with respect to dynamics in~\eqref{sys3}, defined as 
\begin{align}\label{Lie derivative}
	\mathsf{L}\mathcal S_i( x_i)=\partial_{x_i} \mathcal S_i(x_i)(A_i \Theta_i(x_i)x_i + B_iu_i + D_iw_i),
\end{align}
with $\partial_{x_i}  \mathcal S_i(x_i) = \frac{\partial \mathcal S_i(x_i)}{\partial_{x_i}}$.
\end{definition}

\begin{remark}
	A control storage certificate $\mathcal S_i$ captures internal inputs $w_i$,
	resulting from the interaction among subsystems within the interconnection topology. This is apparent from the presence of the quadratic term on the right-hand side of \eqref{eq:martingale2}, referred to as the system's supply rate in dissipativity reasoning \citep{arcak2016networks}, initially served to demonstrate the stability of large-scale networks.
\end{remark}

The order of quantifiers employed in condition~\eqref{eq:martingale2}, \emph{i.e.,} $\forall x_i\in X_i$, $\exists u_i\in U_i$, $\forall w_i\in W_i$, inherently facilitates the development of a \emph{fully-decentralized} controller for $\Xi_i$, mainly since the control input $u_i$ remains independent of the internal input $w_i$ (\emph{i.e.,} states of other subsystems according to \eqref{topology}). Hence, this decentralized structure offers substantial flexibility in the design methodology of controllers for large-scale networks.

\begin{remark}
We impose condition \(\mathds Z_i^{22} \preceq 0\) in Definition~\ref{barrier-condirions-model} to ensure that states of the subsystem do not contribute energy to the CSC \(\mathcal S_i\). In particular, the inequality in condition~\eqref{eq:martingale2} specifically involves the term \(\mathds Z_i^{22}\), which interacts with the state vector \(x_i\). By enforcing \(\mathds Z_i^{22} \preceq 0\), one can ensure that \(x_i^\top \mathds Z_i^{22} x_i \preceq 0\) for all \(x_i\). Hence, this condition is crucial for ensuring that the subsystem’s states maintain a dissipative behavior.
\end{remark}

We now present the following definition to characterize control barrier certificates for interconnected networks \emph{without} internal signals.
\begin{definition} \label{cbc}
	Consider an interconnected network $\mathcal{I}\left(\Xi_1, \ldots, \Xi_N\right)$ as in Definition~\ref{network}, comprising $N$ subsystems $\Xi_i$. Assuming the existence of constants $\lambda,\eta,\mu\in\R^{+}$, with $\mu > \eta$, a function $\mathcal B:X \rightarrow \mathbb{R}^+_0$ is referred to as a control barrier certificate (CBC) for $\Xi$, if
 \begin{subequations}
	\begin{align}\label{sys1-N}
		&\mathcal B(x) \leq \eta,\quad\quad\quad\quad\quad\!\forall x \in X_{0},\\\label{sys2-N}
		&\mathcal B(x) \geq \mu, \quad\quad\quad\quad\quad\!\forall x \in X_{a},
	\end{align} 
	and $\forall x\in X$, $\exists u\in U$, such that
	\begin{align}\label{cbceq}
		\mathsf{L}\mathcal B( x)\mathcal ~\leq - \lambda \mathcal B(x),
	\end{align}
 \end{subequations}
where $\mathsf{L} \mathcal B$ is the Lie derivative of $\mathcal B$
with respect to dynamics in~\eqref{sysN}, defined as 
\begin{align}\label{Lie derivative1}
	\mathsf{L}\mathcal B( x)=\partial_{x} \mathcal B(x)(A(x)x + Bu).
\end{align} Moreover, sets $X_0, X_a \subseteq X$ denote initial and unsafe sets of the interconnected network, respectively. 
\end{definition}
Utilizing Definition~\ref{cbc}, the following theorem, borrowed from~\citep{Pranja}, offers a safety guarantee across the network $\Xi$ within an infinite time horizon.

\begin{theorem}\label{theorem:safe}
Given a network \(\Xi = (A(x), B, X, U)\) as defined in Definition~\ref{network}, assume that \(\mathcal{B}\) qualifies as a CBC for \(\Xi\) according to Definition~\ref{cbc}. Then the network \(\Xi\) is considered safe over an infinite time horizon, implying that all its trajectories originating from \(X_{0} \subseteq X\) never intersect \(X_{a} \subseteq X\) under a controller $u(\cdot)$, e.g.,  \(x_{x_{0,u}}(t) \cap X_a = \emptyset\) for all \(x_0\in X_0\) and for every \(t\in\mathbb{R}^{+}_{0}\).
\end{theorem}

\begin{remark}
In the CBC framework adopted in this paper, the sets of interest are defined by the user based on the desired safety property. The only key requirement is that $X_0 \cap X_a = \emptyset$, i.e., the initial set $X_0$ must not intersect the unsafe set $X_a$. If this condition is violated, the system is unsafe from the outset, rendering any safety analysis meaningless.
\end{remark}

Simply adhering to the criteria set for CSCs does not suffice to ensure the network's safety. Notably, a CSC takes into account the impact of internal inputs $w_i$ via the supply rate in \eqref{eq:martingale2}, capturing the interplay of neighboring subsystems within the interconnection topology. Subsequently, contingent upon certain dissipativity compositional conditions, these CSCs are amalgamated to formulate a CBC for an interconnected network, thereby guaranteeing its safety via Theorem \ref{theorem:safe}. Indeed, designing a CBC alongside its controller to uphold safety in interconnected networks poses a challenging and computationally intensive endeavor, \emph{even when the underlying model is known}. Hence, our compositional strategy involves decomposing this task by concentrating solely on subsystems, initially constructing CSCs and then integrating them under specific conditions to construct a CBC for the interconnected network.

It is apparent that directly satisfying condition~\eqref{eq:martingale2} for CSC is unfeasible, arising from the presence of unknown matrices $A_i$ and $B_i$ within $\mathsf{L}\mathcal{S}_i(x_i)$. Faced with these considerable hurdles, we now delineate the primary problem this work aims to tackle.
\begin{resp}
\begin{problem}\label{problem}~Consider an interconnected network $\mathcal{I}\left(\Xi_1, \ldots, \Xi_N\right)$, comprising $N$ subsystems $\Xi_i$ characterized by unknown matrices $A_i, B_i$. Develop a compositional data-driven approach rooted in dissipativity reasoning to design a CBC and its associated safety controller for the network, thereby ensuring network safety, by constructing CSCs for subsystems through collecting only one noised-corrupted input-state trajectory from each individual subsystem.
\end{problem}
\end{resp}

\section{Data-driven construction of CSCs and local controllers}\label{DD-CBCs}\vspace{-0.2cm}
To address Problem~\ref{problem}, this section outlines our data-driven framework to construct CSCs and their associated local controllers across unknown ct-NPS.
Within our data-driven scheme, we adopt a quadratic form for CSCs, denoted as $\mathcal S_i(x_i) = x_i^\top P_i x_i$, with $P_i \succ 0$. We gather input-state data from unknown ct-NPS during the period $[t_0, t_0 + (\mathcal{T} - 1)\tau]$, with $\mathcal{T}$ denoting the total number of collected samples, while $\tau$ indicates the sampling interval
\begin{subequations}\label{New}
\begin{align}
\mathcal U^{0,\mathcal{T}}_i \!\!&=\! [u_i(t_0)~u_i(t_0 \!+\! \tau)~\dots~u_i(t_0 \!+\! (\mathcal{T} \!-\! 1)\tau)] \label{eq:U0},\\
\mathcal W^{0,\mathcal{T}}_i \!\!&=\! [w_i(t_0)~\!w_i(t_0 \!+\! \tau)~\dots~w_i(t_0 \!+\! (\mathcal{T} \!-\! 1)\tau)] \label{eq:W0},\\
\mathcal X^{0,\mathcal{T}}_i \!\!&=\! [x_i(t_0)~x_i(t_0 \!+\! \tau)~\dots~x_i(t_0 \!+\! (\mathcal{T} \!-\! 1)\tau)] \label{eq:X0},\\
\bar{\mathcal X}^{1,\mathcal{T}}_i \!\!&~\!\!=\! [\dot x_i(t_0)~\dot x_i(t_0\! +\! \tau)~\dots~\dot x_i(t_0 \!+\! (\mathcal{T} \!- 1)\tau)] \label{eq:barX1}.
\end{align}
\end{subequations}
We treat trajectories in \eqref{New}  as a \emph{single input-state trajectory}, collected from each unknown subsystem. Since the state derivatives at the sampling times in \eqref{eq:barX1} cannot be measured directly, we consider them as being corrupted by noise $\Phi_i$, where
\[
\Phi_i \!=\!\bigl[\phi_i(t_0),\,\phi_i(t_0+\tau),\,\dots,\,\phi_i(t_0+(\mathcal{T}-1)\tau)\bigr]
\!\in\! \mathbb{R}^{n_i\times\mathcal{T}}
\]
represent such \emph{unknown-but-bounded} noise. Accordingly, the data we actually record is noisy as $\mathcal{X}_i^{1,\mathcal{T}} = \bar{\mathcal{X}}_i^{1,\mathcal{T}} + \Phi_i$, with \(\Phi_i\) satisfying the following assumption.
\begin{assumption}
The noise matrix \(\Phi_i\) satisfies the following condition
	\begin{equation}\label{noise-bound}
		\Phi_i\,\Phi_i^{\top}\;\preceq\;\Psi_i\,\Psi_i^{\top}\!,
	\end{equation}
where \(\Psi_i\in\mathbb{R}^{n_i\times\mathcal{T}}\) is a known matrix, implying that the energy of the noise remains bounded over the finite time of data collection~\citep{van2020noisy}.
\end{assumption}
In practice, in the case of having a scalar \(\bar{\phi}_i > 0\) such that
\(\|\phi_i(t)\|^2 \le \bar{\phi}_i\) for all \(t \ge 0\), one can set
\begin{equation}\label{noise-bound-2}
\Psi_i\,\Psi_i^{\top}
	\;=\;
\bar{\phi}_i\,\mathcal{T}\,\mathbf{I}_{n_i},
\end{equation}
thus ensuring condition \eqref{noise-bound} is met.
With unknown matrices $A_i$ and $B_i$ embedded within $\mathsf{L}\mathcal{S}_i(x_i)$ in \eqref{Lie derivative}, we now introduce the following lemma, inspired by~\cite{de2019formulas}, aiming to establish a data-driven representation for $A_i\Theta_i(x_i)x_i + B_iu_i$.
\begin{lemma}\label{Lemma1}
	For each subsystem $\Xi_i$, consider a $(\mathcal{T}\times n_i)$  matrix polynomial $ \mathcal Q_i(x_i)$, such that
 \begin{equation}\label{matrixQ}
     {\Theta_i(x_i)} = \mathcal N^{0,\mathcal{T}}_i \mathcal Q_i(x_i),
 \end{equation} 
 with $ \Theta_i(x_i) \in \mathbb{R}^{M_i \times n_i}$ being a state-dependent transformation matrix as in~\eqref{Transformation}
and $\mathcal N^{0,\mathcal{T}}_i$ being an $(M_i\times \mathcal{T})$ full row-rank matrix, derived from the vector $\mathcal R_i(x_i)=\Theta_i(x_i)x_i$~\eqref{Transformation} and samples $\mathcal X^{0,\mathcal{T}}_i$, defined as
\begin{align}\nonumber
		\mathcal N^{0,\mathcal{T}}_i & = \Big[\overbrace{\Theta_i(x_i(t_0))x_i(t_0)}^{\mathcal R_i(x_i(t_0))}~~\overbrace{\Theta_i(x_i(t_0 + \tau))x_i(t_0 + \tau)}^{\mathcal R_i (x_i(t_0 + \tau))}~\dots~\\\label{trjectory-N}& ~~~~\underbrace{\Theta_i (x_i(t_0 + (\mathcal{T} - 1)\tau))(x_i(t_0 + (\mathcal{T} - 1)\tau)}_{\mathcal R_i (x_i(t_0 + (\mathcal{T} - 1)\tau))}\Big].
\end{align}
By designing decentralized controllers $u_i =F_i(x_i) x_i$ with $F_i(x_i) = \mathcal U^{0,\mathcal{T}}_i \mathcal Q_i(x_i)$, the term $A_i\Theta_i(x_i) + B_i F_i(x_i)$ can be equivalently represented in the following data-based form:
 \begin{equation*}
     {A_i\Theta_i(x_i) + B_i F_i(x_i)= (\mathcal X^{1,\mathcal{T}}_i - D_i\mathcal W^{0,\mathcal{T}}_i - \Phi_i) \mathcal Q_i(x_i)}. 
 \end{equation*}
\end{lemma}
{\bf Proof.}
By employing the input-state trajectory specified in~\eqref{New}, one is able to illustrate the data-driven representation of $\dot x_i \!=\!\! A_i\mathcal R_i(x_i) + B_iu_i $ as
\begin{align*} 
	\bar{\mathcal X}^{1,\mathcal{T}}_i &= A_i \mathcal N^{0,\mathcal{T}}_i + B_i\mathcal U^{0,\mathcal{T}}_i + D_i \mathcal W^{0,\mathcal{T}}_i \\
	&= [B_i\quad A_i] \begin{bmatrix}
		\mathcal U^{0,\mathcal{T}}_i\\
		\mathcal N^{0,\mathcal{T}}_i
	\end{bmatrix} + D_i\mathcal W^{0,\mathcal{T}}_i\!\!.
\end{align*}
Accordingly, as $\mathcal{X}_i^{1,\mathcal{T}}:= \bar{\mathcal{X}}_i^{1,\mathcal{T}} + \Phi_i$, one has
\begin{align*} 
	[B_i\quad A_i] \begin{bmatrix}
		\mathcal U^{0,\mathcal{T}}_i\\
		\mathcal N^{0,\mathcal{T}}_i
	\end{bmatrix} = \mathcal X^{1,\mathcal{T}}_i - D_i\mathcal W^{0,\mathcal{T}}_i -\Phi_i.
\end{align*}
\!\!\! According to conditions~\eqref{matrixQ},~\eqref{Transformation} and given the controller matrix $F_i(x_i) = \mathcal U^{0,\mathcal{T}}_i \mathcal Q_i(x_i)$, we have
\begin{align*}
	A_i \mathcal{R}_i\left(x_i\right)+B_i u_i &=A_i\Theta_i(x_i)x_i + B_iu_i\\ &=\left(A_i \Theta_i(x_i) +B_i F_i(x_i) \right) x_i\\ &= [B_i\quad A_i] \begin{bmatrix}
		F_i(x_i)\\
	\Theta_i(x_i)
	\end{bmatrix} x_i\\&= \underbrace{[B_i\quad A_i] \begin{bmatrix}
			\mathcal U^{0,\mathcal{T}}_i\\
			\mathcal N^{0,\mathcal{T}}_i
	\end{bmatrix}}_{\mathcal X^{1,\mathcal{T}}_i - D_i\mathcal W^{0,\mathcal{T}}_i -\Phi_i} \mathcal Q_i(x_i)x_i\\ & = {(\mathcal X^{1,\mathcal{T}}_i - D_i\mathcal W^{0,\mathcal{T}}_i - \Phi_i)} \mathcal Q_i(x_i)x_i.
\end{align*}
Hence, $A_i\Theta_i(x_i) + B_iF_i(x_i)$ can be \emph{equivalently} represented based on it data-driven form as $(\mathcal X^{1,\mathcal{T}}_i - D_i\mathcal W^{0,\mathcal{T}}_i -\Phi_i) \mathcal Q_i(x_i)$. $\hfill\blacksquare$
\begin{remark}
Ensuring that the matrix $\mathcal{N}^{0,\mathcal{T}}_i$ achieves full row-rank requires a minimum sample size $\mathcal{T}$ equal to $M_i+1$ as a sound condition. This criterion is readily verified since $\mathcal{N}^{0,\mathcal{T}}_i$ is derived from sampled data.
\end{remark}

\begin{remark}
If the data is not noisy and $\mathcal{N}_i^{0, \mathcal{T}}$ and $\mathcal{U}_i^{0, \mathcal{T}}$ are both full row rank, then $[B_i ~~ A_i]$ can be identified, allowing one to treat the problem as model-based. However, in our framework, identifying these matrices is challenging due to data noise. Instead,  we only require a data-driven representation of the closed-loop subsystem term $A_i\Theta_i(x_i)x_i + B_i u_i$, assuming that only $\mathcal{N}_i^{0, \mathcal{T}}$ is full row rank, rather than both $\mathcal{N}_i^{0, \mathcal{T}}$ and $\mathcal{U}_i^{0, \mathcal{T}}$\!\!.
\end{remark}

By employing the data-based representation of $A_i\Theta_i(x_i) + B_iF_i(x_i)$  in Lemma~\ref{Lemma1}, we propose the following theorem, as one of the main findings of our work, to construct CSCs from noisy data while synthesizing decentralized controllers for unknown subsystems.

\begin{theorem}\label{Thm:main2}
	Consider an unknown ct-NPS $\Xi_i$ as in \eqref{sys2}, with its data-based representation $A_i\Theta_i(x_i) + B_iF_i(x_i) = (\mathcal X^{1,\mathcal{T}}_i - D_i\mathcal W^{0,\mathcal{T}}_i-\Phi_i) \mathcal Q_i(x_i)$ as per Lemma \ref{Lemma1}. Suppose there exists a matrix polynomial
	$\mathcal H_i(x_i) \in \mathbb R^{\mathcal{T} {\times n_i}}$ such that
		\begin{align}\label{matrixP}
		& \mathcal N_i^{0,\mathcal{T}}\mathcal H_i(x_i) = {\Theta_i(x_i)}P_i^{-1}, \:\text{with} ~ P_i \succ 0.
	\end{align}
    Let there exist constants $\eta_i,\mu_i, \alpha_i \in\R^{+}$\!, and matrices $\bar{\mathds Z}^{11}_i, \bar{\mathds Z}^{12}_i, \bar{\mathds Z}^{21}_i, \bar{\mathds Z}^{22}_i$ of appropriate dimensions, with $\bar{\mathds Z}^{21}_i = \bar{\mathds Z}^{{12}^\top}_i\!\!,$ and $ \bar{\mathds Z}^{22}_i \preceq 0$, so that
    \begin{subequations}\label{data-cons}
	\begin{align}\label{con1a}
	&x_i^\top [\Theta^\dagger_i(x_i)\mathcal N^{0,\mathcal{T}}_i\mathcal H_i(x_i)]^{-1} x_i \leq \eta_i,\quad\forall x_i \in X_{0_i},\\\label{con2a}
	& x_i^\top [\Theta^\dagger_i(x_i)\mathcal N^{0,\mathcal{T}}_i\mathcal H_i(x_i)]^{-1} x_i  \geq \mu_i,\quad\forall x_i \in X_{a_i},
		\end{align}
and $\forall x_i \in X_{i}$,
		\begin{align}\label{con3a}
	{\begin{bmatrix}
	- \mathcal{G}_i	+ \bar{\mathds Z}^{22}_i &&& \mathcal{H}^\top_i(x_i)&&& \bar{\mathds Z}^{21}_i\\
	\mathcal{H}_i(x_i)&&& \alpha_i \mathbf{I}_{\mathcal{T}}&&& \mathbf{0}_{\mathcal{T} \times n_i }\\
	\bar{\mathds Z}^{12}_i&&&\mathbf{0}_{n_i \times \mathcal{T}}&&&\bar{\mathds Z}^{11}_i -\frac{1}{\pi_i} D_i^\top D_i
	\end{bmatrix}\succeq 0,}
	\end{align}
 \end{subequations}
with 
\begin{align}\notag
 \mathcal{G}_i &= (\mathcal{X}_i^{1,\mathcal{T}} - D_i\,\mathcal{W}_i^{0,\mathcal{T}})\,
\mathcal{H}_i(x_i)
+ \alpha_i\,\Psi_i\Psi_i^\top\\\notag &~~~
+ \mathcal{H}_i(x_i)^\top\,
(\mathcal{X}_i^{1,\mathcal{T}} - D_i\,\mathcal{W}_i^{0,\mathcal{T}})^\top
+ \pi_i\,\mathbf{I}_{n_i}\\\label{Mathcal-G} &~~~ + \lambda_i \Theta^{\dagger}_i(x_i)\mathcal N^{0,\mathcal{T}}_i\mathcal H_i(x_i),
\end{align}
for some $\lambda_i,\pi_i \in\R^{+}$. Then, the function
\[
\mathcal{S}_i(x_i) = x_i^\top {[\Theta_i^\dagger(x_i)\mathcal{N}_i^{0,\mathcal{T}}\mathcal{H}_i(x_i)]^{-1}} x_i,
\]
is a CSC for subsystem $\Xi_i$, and the corresponding decentralized controller is designed as
\[
u_i = \mathcal{U}_i^{0,\mathcal{T}} \mathcal{H}_i(x_i){[\Theta_i^\dagger(x_i)\mathcal{N}_i^{0,\mathcal{T}}\mathcal{H}_i(x_i)]^{-1}}x_i.
\]
Furthermore, the matrices associated with this CSC and its decentralized controller are specified by $\mathds{Z}^{11}_i = \bar{\mathds{Z}}^{11}_i, \mathds{Z}^{12}_i = \bar{\mathds{Z}}^{12}_i[\Theta_i^\dagger(x_i)\mathcal{N}_i^{0,\mathcal{T}}\mathcal{H}_i(x_i)]^{-1}, \mathds{Z}^{21}_i = \mathds{Z}^{12^\top}_i, \mathds{Z}^{22}_i \\= [\Theta_i^\dagger(x_i)\mathcal{N}_i^{0,\mathcal{T}}\mathcal{H}_i(x_i)]^{-1}\bar{\mathds{Z}}^{22}_i[\Theta_i^\dagger(x_i)\mathcal{N}_i^{0,\mathcal{T}}\mathcal{H}_i(x_i)]^{-1}.$
\end{theorem}

{\bf Proof.} Given that $\mathcal{S}_i(x_i)\!=\! x_i^\top \![\Theta^\dagger_i(x_i)\mathcal N^{0,\mathcal{T}}_i\!\mathcal H_i(x_i)]^{-1} x_i$ and $\mathcal N_i^{0,\mathcal{T}}\mathcal H_i(x_i) =\Theta_i(x_i) P_i^{-1}$ according to \eqref{matrixP}, it is clear that fulfilling~\eqref{con1a} and~\eqref{con2a} leads to the implications of conditions \eqref{subsys1} and \eqref{subsys2}, respectively.\\
Next, we proceed with showing condition~\eqref{eq:martingale2}, as well. Since $\Theta_i(x_i)P_i^{-1} =  \mathcal N^{0,\mathcal{T}}_i\mathcal H_i(x_i)$ as per \eqref{matrixP}, and $P_i^{-1}P_i = \mathbf{I}_{n_i}$, then $ \Theta_i(x_i)= \mathcal N^{0,\mathcal{T}}_i\mathcal H_i(x_i)P_i$. Since $\Theta_i(x_i) = \mathcal N^{0,\mathcal{T}}_i \mathcal Q_i(x_i)$ according to \eqref{matrixQ},  then one can set $\mathcal Q_i(x_i) = \mathcal H_i(x_i)P_i$ and, accordingly, $\mathcal Q_i(x_i) P_i^{-1} = \mathcal H_i(x_i)$. Given that $A_i\Theta_i(x_i) + B_iF_i(x_i) = (\mathcal X^{1,\mathcal{T}}_i - D_i\mathcal W^{0,\mathcal{T}}_i -\Phi_i) \mathcal Q_i(x_i)$ according to Lemma \ref{Lemma1}, one has
\begin{align}\nonumber
	(A_i &\Theta(x_i)+ B_iF_i(x_i))P_i^{-1} \\\nonumber &= (\mathcal X^{1,\mathcal{T}}_i - D_i\mathcal W^{0,\mathcal{T}}_i -\Phi_i) \underbrace{\mathcal Q_i(x_i)P_i^{-1}}_{\mathcal H_i(x_i)}\\\label{close-loop} &=(\mathcal X^{1,\mathcal{T}}_i - D_i\mathcal W^{0,\mathcal{T}}_i-\Phi_i)\mathcal H_i(x_i).
\end{align}
By employing the definition of Lie derivative in~\eqref{Lie derivative}, we have 
\begin{align*}
	\mathsf{L}\mathcal S_i(x_i) &=\partial_{x_i} \mathcal S_i(x_i)(A_i\mathcal R_i(x_i) + B_i u_i + D_iw_i)
	\\ &=2 x_i^\top P_i \big(\!(A_i \Theta_i(x_i) \!+\!B_i\mathcal{F}_i(x_i)) x_i\!+\! D_i w_i\big) 
	\\ &=2 x_i^\top P_i \big(A_i \Theta_i(x_i) \!+\!B_i F_i(x_i)\big) x_i  + 2 \underbrace{x_i^\top P_i}_{a_i} \underbrace{D_i w_i}_{b_i}.
\end{align*}
By applying the Cauchy-Schwarz inequality~\citep{bhatia1995cauchy}, namely $a_i^\top b_i \leq \|a_i\| \|b_i\|$ for any $a_i^\top, b_i \in \mathbb{R}^{n_i}$, and subsequently using Young's inequality~\citep{young1912classes}, given by $\|a_i\| \|b_i\| \leq \frac{\pi_i}{2} \|a_i\|^2 + \frac{1}{2\pi_i} \|b_i\|^2$ for any $\pi_i \in \mathbb{R}^+$, it follows that
\begin{align}\notag
	\mathsf{L}\mathcal S_i(x_i) &\leq  2x_i^\top P_i \big(A_i\Theta_i(x_i)+B_iF_i(x_i)\big) x_i\\ \label{new5}
	&~~~+ \pi_i x_i^\top P_i P_i x_i x_i + \frac{1}{\pi_i} w_i^\top D_i^\top D_i w_i.
\end{align}
By expanding the expression given above and factorizing the term $ x_i^\top P_i$ from left and $P_i x_i$ from right, the inequality in~\eqref{new5} can be equivalently represented as
\begin{align*}
	\mathsf{L} \mathcal S_i(x_i) &\leq  x_i^\top P_i\Big[(A_i\Theta_i(x_i) + B_i F_i(x_i))P_i^{-1}\\ 
	& ~~~+ P_i^{-1} ( A_i \Theta_i(x_i)\!+\! B_i F_i(x_i))^\top {+\pi_i \mathbf{I}_{n_i}} \Big]\! P_i x_i \\ &~~~ +\frac{1}{\pi_i} w_i^\top D_i^\top D_i w_i.
\end{align*}
Then, by employing~\eqref{close-loop}, one can replace $(A_i\Theta_i(x_i) + B_i F_i(x_i))P_i^{-1}$ with its data-based representation and obtain
\begin{align}\notag
	\mathsf{L} \mathcal S_i(x_i) &\leq  x_i^\top \!\!P_i\Big[ (\mathcal X^{1,\mathcal{T}}_i \!\!\!-\! D_i\mathcal W^{0,\mathcal{T}}_i -\Phi_i)\mathcal H_i(x_i) \\ \notag
	&~~~+\! \mathcal H_i(x_i)^\top (\mathcal X^{1,\mathcal{T}}_i \!\!\!-\! D_i\mathcal W^{0,\mathcal{T}}_i -\Phi_i)^\top + \pi_i \mathbf{I}_{n_i} \Big] P_i x_i\\ \label{Young-ex} &~~~+\frac{1}{\pi_i} w_i^\top D_i^\top D_i w_i.
\end{align}
By applying the Young's inequality for matrices~\citep{Ando1995}, which states that 
$\mathbb{P}_i(x_i) \mathbb{Q}_i(x_i)^\top + \mathbb{Q}_i(x_i) \mathbb{P}_i(x_i)^\top \geq -\alpha_i \mathbb{P}_i(x_i) \mathbb{P}_i(x_i)^\top - \alpha^{-1}_i \mathbb{Q}_i(x_i) \mathbb{Q}_i(x_i)^\top$ 
for any scalar $\alpha_i > 0$, to 
$- \Phi_i\mathcal{H}_i(x_i) - \mathcal{H}_i(x_i)^\top\Phi_i^\top$, we have
\begin{align*}
	\mathsf{L} \mathcal S_i(x_i) &\leq  x_i^\top \!\!P_i\Big[(\mathcal X_i^{1,\mathcal{T}} -D_i \mathcal W_i^{0,\mathcal{T}} \!)\mathcal H_i(x_i) + {\alpha_i \Phi_i \Phi_i^\top}\!\\ &~~~+ \frac{1}{\alpha_i}\mathcal{H}_i(x_i)^\top\mathcal{H}_i(x_i)\!+\!\mathcal H_i(x_i)^\top\!(\mathcal X_i^{1,\mathcal{T}} \!\!\!-\!\! D_i \mathcal W_i^{0,\mathcal{T}})^\top
	\\ &~~~+ \pi_i \mathbf{I}_{n_i} \Big]P_i x_i +\frac{1}{\pi_i} w_i^\top D_i^\top D_i w_i.\\
\end{align*}
Then, according to the inequality in~\eqref{noise-bound}, one has
\begin{align*}
	\mathsf{L}\mathcal S_i(x_i) &\leq  x_i^\top \!\!P_i\Big[(\mathcal X_i^{1,\mathcal{T}} -D_i \mathcal W_i^{0,\mathcal{T}} \!)\mathcal H_i(x_i) + {\alpha_i \Psi_i \Psi_i^\top}\!\\ &~~~+ \frac{1}{\alpha_i}\mathcal{H}_i(x_i)^\top\mathcal{H}_i(x_i)\!+\!\mathcal H_i(x_i)^\top\!(\mathcal X_i^{1,\mathcal{T}} \!\!\!-\!\! D_i \mathcal W_i^{0,\mathcal{T}})^\top	\\ &~~~+ \pi_i \mathbf{I}_{n_i} \Big]P_i x_i +\frac{1}{\pi_i} w_i^\top D_i^\top D_i w_i\\
		&= \begin{bmatrix}
		P_i x_i\\		
		w_i
	\end{bmatrix}^\top\!\begin{bmatrix}
		\mathds{S}_i   & \mathbf{0}_{n_i \times n_i}\\
		\mathbf{0}_{n_i \times n_i} & \frac{1}{\pi_i} D_i^\top D_i
	\end{bmatrix} \begin{bmatrix}
		P_i x_i\\
		w_i
	\end{bmatrix}\!\!,
\end{align*}
with $ \mathds{S}_i = (\mathcal{X}_i^{1,\mathcal{T}} - D_i\,\mathcal{W}_i^{0,\mathcal{T}})\,
\mathcal{H}_i(x_i)
+ {\alpha_i\,\Psi_i\Psi_i^\top
+ \frac{1}{\alpha_i}\,\mathcal{H}_i(x_i)^\top}\\\,\mathcal{H}_i(x_i)
+ \mathcal{H}_i(x_i)^\top\,
(\mathcal{X}_i^{1,\mathcal{T}} - D_i\,\mathcal{W}_i^{0,\mathcal{T}})^\top
+ \pi_i\,\mathbf{I}_{n_i}$. Since, according to the Schur complement~\citep{zhang2006schur}, we can conclude that the satisfaction of condition~\eqref{con3a} results in
\begin{align*}
\begin{bmatrix}
	\mathds{S}_i   & \mathbf{0}_{n_i \times n_i}\\
	\mathbf{0}_{n_i \times n_i} & \frac{1}{\pi_i} D_i^\top D_i
\end{bmatrix} \!\! \preceq\!\! \begin{bmatrix}
- \lambda_i \Theta^\dagger_i(x_i)\mathcal N^{0,\mathcal{T}}_i\mathcal H_i(x_i)+ \bar{\mathds Z}^{22}_i & \bar{\mathds Z}^{21}_i\\
\bar{\mathds Z}^{12}_i&\bar{\mathds Z}^{11}_i
\end{bmatrix}\!\!,
\end{align*}	
then we have
\begin{align*}
	\mathsf{L}\mathcal S_i(x_i) &\leq \begin{bmatrix}
		P_i x_i\\		
		w_i
	\end{bmatrix}^\top\!\!\! \begin{bmatrix}
		- \lambda_i \Theta^\dagger_i(x_i)\mathcal N^{0,\mathcal{T}}_i\mathcal H_i(x_i)+ \bar{\mathds Z}^{22}_i & \bar{\mathds Z}^{21}_i\\
		\bar{\mathds Z}^{12}_i&\bar{\mathds Z}^{11}_i
	\end{bmatrix} \!\!\times\\ 
	&~~~~ \begin{bmatrix}
		P_i x_i \\		
		w_i
	\end{bmatrix} \\ & = - \lambda_i x_i^\top P_i \overbrace{\underbrace{[\Theta^\dagger_i(x_i)\mathcal N_i^{0,T}\mathcal H_i(x_i)]}_{P_i^{-1}} P_i}^{\mathbf I_{n_i}} x_i\\ 
	& ~~~+ \begin{bmatrix}
		x_i\\		
		w_i
	\end{bmatrix}^\top\! \begin{bmatrix}
		P_i\bar{\mathds Z}^{22}_iP_i & P_i\bar{\mathds Z}^{21}_i\\
		\bar{\mathds Z}^{12}_iP_i&\bar{\mathds Z}^{11}_i
	\end{bmatrix} \begin{bmatrix}
		x_i\\		
		w_i
	\end{bmatrix}\\
	&= - \lambda_i x_i^\top P_i x_i + \begin{bmatrix}
		 x_i \\		
		w_i
	\end{bmatrix}^\top\! \begin{bmatrix}
		P_i\bar{\mathds Z}^{22}_iP_i &&& P_i\bar{\mathds Z}^{21}_i\\
		\bar{\mathds Z}^{12}_iP_i&&&\bar{\mathds Z}^{11}_i
	\end{bmatrix} \begin{bmatrix}
		x_i\\		
		w_i
	\end{bmatrix}\!\!.
\end{align*}
\!\!\! By replacing $P_i$ with $\mathcal [ \Theta^\dagger_i(x_i\mathcal N^{0,\mathcal{T}}_i\mathcal H_i(x_i)]^{-1}$ according to~\eqref{matrixP}, we obtain the following
\begin{align*}
	& \mathsf{L}\mathcal S_i(x_i) \leq - \lambda_i \mathcal S_i(x_i) + \begin{bmatrix}
		w_i\\	
		x_i
	\end{bmatrix}^\top\! \begin{bmatrix}
		\mathds Z^{11}_i && \mathds Z^{12}_i\\
		\mathds Z^{21}_i&&\mathds Z^{22}_i
	\end{bmatrix} \begin{bmatrix}
		w_i\\	
		x_i
	\end{bmatrix}\!\!,
\end{align*}
with 
\begin{align*}
        \mathds{Z}^{11}_i &= \bar{\mathds{Z}}^{11}_i\!\!,
        ~~~\mathds{Z}^{12}_i = \bar{\mathds{Z}}^{12}_i[\Theta_i^\dagger(x_i)\mathcal{N}_i^{0,\mathcal{T}}\mathcal{H}_i(x_i)]^{-1}\!\!, \mathds{Z}^{21}_i \!\!= \!\!\mathds{Z}^{12^\top}_i\!\!\!\!,\\
		\mathds{Z}^{22}_i  &= [\Theta_i^\dagger(x_i)\mathcal{N}_i^{0,\mathcal{T}}\mathcal{H}_i(x_i)]^{-1}\bar{\mathds{Z}}^{22}_i[\Theta_i^\dagger(x_i)\mathcal{N}_i^{0,\mathcal{T}}\mathcal{H}_i(x_i)]^{-1}\!\!,
\end{align*}
satisfying condition~\eqref{eq:martingale2}. Hence, the function $\mathcal{S}_i(x_i) = x_i^\top [\Theta_i^\dagger(x_i)\mathcal{N}_i^{0,\mathcal{T}}\mathcal{H}_i(x_i)]^{-1} x_i$ 
is a CSC for subsystem $\Xi_i$ and its corresponding \emph{decentralized} controller is $u_i = \mathcal{U}_i^{0,\mathcal{T}} \underbrace{\mathcal{H}_i(x_i)[\Theta_i^\dagger(x_i)\mathcal{N}_i^{0,\mathcal{T}}\mathcal{H}_i(x_i)]^{-1}}_{\mathcal{Q}_i(x_i)}x_i.$ $\hfill\blacksquare$

\begin{remark}
In order to handle the noisy data, condition~\eqref{con3a} explicitly incorporates contributions from both \( \mathcal H_i(x_i) \) and \( \mathcal H^\top_i(x_i) \). This adjustment results in an increased number of decision variables corresponding to the horizon of collected data, which is expected in the noisy-data setting.
\end{remark}

\subsection{Computation of CSCs and local controllers}\label{sos_sedumi}\vspace{-0.2cm}
The set of conditions in~\eqref{data-cons} can be reformulated as a sum-of-squares (SOS) optimization program and implemented using available software tools such as \textsf{SOSTOOLS} \citep{SOSTOOLS}, coupled with semi-definite programming (SDP) solvers like \textsf{Mosek} \citep{mosek}. We provide the subsequent lemma to outline such an SOS formulation, aiming to construct a CSC and its controller for each unknown subsystem.

\begin{lemma}\label{SOS}
 	Let sets $X_i$, $X_{0_i}$, and $X_{a_i}$ each be defined by vectors of polynomial inequalities in the form of $X_i = \{x_i \in \mathbb{R}^{n_i} | b_i(x_i) \geq 0\}$, $X_{0_i} = \{x_i \in \mathbb{R}^{n_i} | b_{0_i}(x_i) \geq 0\}$, and $X_{a_i} = \{x_i \in \mathbb{R}^{n_i} | b_{a_i}(x_i) \geq 0\}$, respectively. Suppose there exist a matrix polynomial $\mathcal H_i(x_i)$, constants $\lambda_i,\mu_i, \eta_i, \pi_i \in \mathbb{R}^{+}$, matrices $\bar{\mathds Z}^{11}_i, \bar{\mathds Z}^{12}_i, \bar{\mathds Z}^{21}_i, \bar{\mathds Z}^{22}_i$ of appropriate dimensions, with $\bar{\mathds Z}^{21}_i = \bar{\mathds Z}^{{12}^\top}_i\!\!,$ and $ \bar{\mathds Z}^{22}_i \preceq 0$, and vectors of sum-of-squares polynomials $\varphi_{0_i}(x_i)$, $\varphi_{a_i}(x_i)$, $\varphi_i(x_i)$, such that
 	\begin{subequations}\label{SOS6}
 		\begin{align}\label{SOS0}
 			& -x_i^\top [\Theta_i^\dagger(x_i)\mathcal N^{0,\mathcal{T}}_i\mathcal H_i(x_i)]^{-1} x_i\!-\!\varphi_{0_i}^\top\left(x_i\right) b_{0_i}(x_i){+\eta_i},\\\label{SOS1}
 			& \quad \: x_i^\top [\Theta_i^\dagger(x_i)\mathcal N^{0,\mathcal{T}}_i\mathcal H_i(x_i)]^{-1}  x_i\!-\!\varphi_{a_i}^\top\left(x_i\right) b_{a_i}(x_i){-\mu_i} \text {, } \\\notag
 			& {\begin{bmatrix}
 				- \mathcal{G}_i	 + \bar{\mathds Z}^{22}_i &&& \mathcal{H}^\top_i(x_i)&&& \bar{\mathds Z}^{21}_i\\
 				\mathcal{H}_i(x_i)&&& \alpha_i \mathbf{I}_{\mathcal{T}}&&& \mathbf{0}_{\mathcal{T} \times n_i }\\
 				\bar{\mathds Z}^{12}_i&&&\mathbf{0}_{n_i \times \mathcal{T}}&&&\bar{\mathds Z}^{11}_i -\frac{1}{\pi_i} D_i^\top D_i
 			\end{bmatrix}}\\\label{SOS2} 
 			&~~~-(\varphi_i^\top\left(x_i\right) b_i(x_i))\mathbf{I}_{(2n_i + \mathcal{T})},
 		\end{align}
 	\end{subequations}
 	are SOS polynomials with $\mathcal{G}_i$ as in~\eqref{Mathcal-G}. Then $\mathcal S_i(x_i) =  x_i^\top {[\Theta_i^\dagger(x_i)\mathcal N^{0,\mathcal{T}}_i\mathcal H_i(x_i)]^{-1}} x_i$ is a CSC satisfying conditions in~\eqref{data-cons} with $u_i \!\!=\! {\mathcal U^{0,\mathcal{T}}_i \!\!\mathcal H_i(x_i)[\Theta_i^\dagger(x_i)\mathcal N^{0,\mathcal{T}}_i\!\!\mathcal H_i(x_i)]^{-1}} x_i$ being  its corresponding controller.
 \end{lemma}

{\bf Proof.} Since $\varphi_{0_i}(x_i)$ in \eqref{SOS0} is an SOS polynomial, it follows that $\varphi_{0_i}^\top(x_i) b_{0_i}(x_i) \geq 0$ within $X_{0_i}=\{x_i \in \mathbb{R}^{n_i} | b_{0_i}(x_i) \!\geq\! 0\}$. Given that $x_i[\Theta_i^\dagger(x_i)\mathcal N^{0,\mathcal{T}}_i\!\mathcal H_i(x_i)]^{-1} x_i$ is also an SOS polynomial (since $[\Theta_i^\dagger(x_i)\mathcal N^{0,\mathcal{T}}_i\mathcal H_i(x_i)]^{-1}  = P_i$ is positive definite) and thus non-negative, fulfillment of \eqref{SOS0} ensures condition \eqref{con1a}. Similarly, satisfying condition \eqref{SOS1} implies the fulfillment of condition \eqref{con2a}. We now proceed with showing condition~\eqref{con3a}, as well. Given that $\varphi_{i}(x_i)$ in~\eqref{SOS2} is an SOS polynomial, it implies that $\varphi_{i}^\top(x_i) b_{i}(x_i) \geq 0$ within  $X_{i}=\{x_i \in \mathbb{R}^{n_i} | b_{i}(x_i) \geq 0\}$. Since~\eqref{SOS2} is also an SOS polynomial, it entails
\begin{align*}
&\begin{bmatrix}
	- \mathcal{G}_i	+ \bar{\mathds Z}^{22}_i &&& \mathcal{H}^\top_i(x_i)&&& \bar{\mathds Z}^{21}_i\\
	\mathcal{H}_i(x_i)&&& \alpha_i \mathbf{I}_{\mathcal{T}}&&& \mathbf{0}_{\mathcal{T} \times n_i }\\
	\bar{\mathds Z}^{12}_i&&&\mathbf{0}_{n_i \times \mathcal{T}}&&&\bar{\mathds Z}^{11}_i -\frac{1}{\pi_i} D_i^\top D_i
\end{bmatrix}\\
&~~~-(\varphi_i^\top\left(x_i\right) b_i(x_i))\mathbf{I}_{(2n_i + \mathcal{T})} \succeq 0,
\end{align*}
\!\!\! concluding that satisfaction of \eqref{SOS2} ensures the fulfillment of condition \eqref{con3a}. $\hfill\blacksquare$
 \vspace{-0.01cm}
 
\begin{remark}
The SOS optimization programs generally face scalability challenges, which can cause problems when coping with relatively high-dimensional subsystems. Despite this, our framework has alleviated the scalability difficulties. In fact, our matrix transformation allows us to define CSC as $\mathcal{S}_i(x_i) = x_i^\top P_i x_i$. Accordingly, the corresponding matrices depend on the number of subsystem's state variables $n_i$, i.e., $\mathcal{Q}_i(x_i) \in \mathbb{R}^{\mathcal{T} \times n_i}$ and $\mathcal{H}_i(x_i) \in \mathbb{R}^{\mathcal{T} \times n_i}$, which is most of the time lower than $M_i$ (cf. all case studies). Consequently, the number of decision variables in our work is lower, resulting in mitigating scalability challenges. Nevertheless, we acknowledge that our data-driven approach may still encounter some scalability issues when either the subsystems are relatively high dimensional, or the matrix $\Theta_i(x_i)$ involves a high degree of monomials.
 \end{remark}
 
\section{Compositional construction of CBC for interconnected network}\label{Compositional}
Here, we provide a compositional framework based on \emph{dissipativity reasoning} for constructing a CBC for an interconnected network based on CSCs of its individual subsystems, derived from noisy data in Theorem \ref{Thm:main2}. To do so, we offer the following theorem.

\begin{theorem}\label{dissipativity}
Consider an interconnected network $\Xi=\mathcal{I}\left(\Xi_1, \ldots, \Xi_N\right)$, composed of $N$ ct-NPS $\Xi_i, i \in\{1, \ldots, N\}$, with an interconnection topology $\mathds M$ as in \eqref{topology}. Assume each subsystem $\Xi_i$ admits a CSC $ \mathcal{S}_i$ based on noisy data according to Theorem \ref{Thm:main2}. If
 \begin{subequations}
\begin{align}\label{dissi}
& \sum_{i=1}^N \eta_i< \sum_{i=1}^N \mu_i, \\
& {\left[\begin{array}{c}
\mathds M \\
\mathbf I_{n}
\end{array}\right]^\top \mathds Z^{\text {comp }}\left[\begin{array}{c}
\mathds M \\\label{LMI}
\mathbf I_{n}
\end{array}\right] \preceq 0,}
\end{align}
 \end{subequations}
with $n = \sum_{i=1}^N n_i$, and 
\begin{align}\label{compose2}
	\mathds Z^{\text {comp }}:=\left[\begin{array}{llllll}
		\mathds Z_1^{11} & & & \mathds Z_1^{12} & & \\
		& \ddots & & & \ddots & \\
		& & \mathds Z_N^{11} & & & \mathds Z_N^{12} \\
		\mathds Z_1^{21} & & & \mathds Z_1^{22} & & \\
		& \ddots & & & \ddots & \\
		& & \mathds Z_N^{21} & & & \mathds Z_N^{22}
	\end{array}\right]\!\!,
\end{align}
then
\begin{align}\label{compose1}
\mathcal{B}(x)=\sum_{i=1}^N \mathcal{S}_i (x_i)= \sum_{i=1}^N x_i^\top {[\Theta_i^\dagger(x_i)\mathcal N^{0,\mathcal{T}}_i\mathcal H_i(x_i)]^{-1}} x_i,
\end{align}
is a CBC for the interconnected network $\Xi$ with $\lambda=\underset{i}{\min}\{\lambda_i\}$, $\eta=\sum_{i=1}^N \eta_i$, and $\mu=\sum_{i=1}^N \mu_i$. Additionally, $u=[u_1; \cdots; u_N]$ with 
\begin{equation}\label{local-controllers}
	u_i = \mathcal U^{0,\mathcal{T}}_i \mathcal H_i(x_i){[\Theta_i^\dagger(x_i)\mathcal N^{0,\mathcal{T}}_i\mathcal H_i(x_i)]^{-1}} x_i,
\end{equation}
$i\in\{1,\dots, N\}$ is its safety controller. 
\end{theorem}

{\bf Proof.} We first show that $\mathcal{B}(x)$ as~\eqref{compose1} satisfies conditions~\eqref{sys1-N} and~\eqref{sys2-N}. For any $x:=\left[x_1 ; \ldots ; x_N\right] \in$ $X_0=\prod_{i=1}^N X_{0_i}$ and from~\eqref{subsys1}, we have
\begin{equation*}
	\mathcal{B}(x)=\sum_{i=1}^N \mathcal{S}_i\left(x_i\right) \leq \sum_{i=1}^N \eta_i=\eta.
\end{equation*}
Similarly for any $x:=\left[x_1 ; \ldots ; x_N\right] \in X_a=\prod_{i=1}^N X_{a_i}$ and from~\eqref{subsys2}, one has
\begin{equation*}
	\mathcal{B}(x)=\sum_{i=1}^N \mathcal{S}_i\left(x_i\right) \geq \sum_{i=1}^N \mu_i=\mu,
\end{equation*}
fulfilling conditions \eqref{sys1-N} and \eqref{sys2-N}, with $\eta=\sum_{i=1}^N \eta_i$ and $\mu=\sum_{i=1}^N \mu_i$. Since $\sum_{i=1}^N \eta_i<\sum_{i=1}^N \mu_i$ according to \eqref{dissi}, one has $\mu>\eta$.

We now show that $\mathcal{B}(x)$ satisfies condition \eqref{cbceq}, as well. By employing the interconnection constraint \eqref{topology} and the compositional condition \eqref{LMI}, one can reach the following chain of inequalities:
\begin{equation*}
	\begin{aligned}
		\mathsf{L}\mathcal B(x)&= \mathsf{L}\sum_{i=1}^{N} \mathcal S_i(x_i) =\sum_{i=1}^{N}  \mathsf{L} \mathcal S_i(x_i) \leq -\sum_{i=1}^{N} \lambda_i \mathcal S_i(x_i)\\ &+ \sum_{i=1}^N\begin{bmatrix}
			w_i \\
			x_i
		\end{bmatrix}^\top {\begin{bmatrix}
			\mathds Z_i^{11} & \mathds Z_i^{12} \\
			\mathds Z_i^{21} & \mathds Z_i^{22}
		\end{bmatrix}}\begin{bmatrix}
			w_i \\
			x_i
		\end{bmatrix}\\ &= -\sum_{i=1}^{N} \lambda_i \mathcal S_i(x_i)\\ 
		&~~~+\begin{bmatrix}
			w_1\\
			\vdots\\
			w_N\\
			x_1\\
			\vdots\\
			x_N
		\end{bmatrix}^\top {\begin{bmatrix}
			\mathds Z_1^{11} & & & \mathds Z_1^{12} & & \\
			& \ddots & & & \ddots & \\
			& & \mathds Z_N^{11} & & & \mathds Z_N^{12} \\
			\mathds Z_1^{21} & & & \mathds Z_1^{22} & & \\
			& \ddots & & & \ddots & \\
			& & \mathds Z_N^{21} & & & \mathds Z_N^{22}
		\end{bmatrix}}\!\!\begin{bmatrix}
			w_1\\
			\vdots\\
			w_N\\
			x_1\\
			\vdots\\
			x_N
		\end{bmatrix}\\
            \end{aligned}
	\end{equation*}
\begin{equation*}
\begin{aligned}
		&=-\sum_{i=1}^{N} \lambda_i \mathcal S_i(x_i)\\ 
		&~~~+\begin{bmatrix}
			\mathds M\begin{bmatrix}
				x_1 \\
				\vdots \\
				x_N
			\end{bmatrix} \\
			x_1 \\
			\vdots \\
			x_N
		\end{bmatrix}^\top \mathds Z^{comp}\begin{bmatrix}
			\mathds M\begin{bmatrix}
				x_1 \\
				\vdots \\
				x_N
			\end{bmatrix} \\
			x_1 \\
			\vdots \\
			x_N
		\end{bmatrix}\\ &=-\sum_{i=1}^{N} \lambda_i \mathcal S_i(x_i)\\ 
		&~~~+\begin{bmatrix}
			x_1 \\
			\vdots \\
			x_N
		\end{bmatrix}^\top \begin{bmatrix}
			\mathds M \\
			\mathbf I_{n}
		\end{bmatrix}^\top \mathds Z^{comp}\begin{bmatrix}
			\mathds M \\
			\mathbf I_{n}
		\end{bmatrix}\begin{bmatrix}
			x_1 \\
			\vdots \\
			x_N
		\end{bmatrix} \\ &\leq -\sum_{i=1}^{N} \lambda_i \mathcal S_i(x_i) = - \lambda \mathcal B(x),
	\end{aligned}
\end{equation*}
where $\lambda=\underset{i}{\min}\{\lambda_i\}$, $i \in\{1, \ldots, N\}$. Then $\mathcal B(x)$ as in \eqref{compose1} is a CBC  for the interconnected network $\Xi$ with its safety controller as $u=[u_1; \cdots; u_N]$, where $u_i$ as~in~\eqref{local-controllers}, $i\in\{1,\dots, N\}$.$\hfill\blacksquare$

\begin{algorithm}[t!]
	\caption{Compositional data-driven design of CBC and its safety controller}\label{Alg:1}
	\begin{algorithmic}[1]
			\REQUIRE Sets $X_{0_i}, X_{a_i}$, a dictionary of $\mathcal R_i(x_i)$ up to a certain degree,  matrix $\Psi_i$
			\FOR{$i = 1, \cdots, N$}\label{Step1}
			\STATE Gather input-state trajectories $\mathcal{U}_i^{0,\mathcal{T}}$, $\mathcal{W}_i^{0,\mathcal{T}}$, $\mathcal{X}_i^{0,\mathcal{T}}$, $\mathcal X^{1,\mathcal{T}}_i$ as specified in~\eqref{New} 
			\STATE Construct $\mathcal{N}^{0,\mathcal{T}}_i$ as in~\eqref{trjectory-N}
			\STATE Select a-priori constants $\lambda_i$ and $\pi_i$ in~\eqref{SOS2}
			\STATE\label{Step5} Use \textsf{SOSTOOLS} to enforce conditions~\eqref{matrixP}\footnotemark[1] and~\eqref{SOS2}, thus designing $\mathcal{H}_i(x)$, CSCs $\mathcal{S}_i(x_i) =x_i^\top [\Theta_i^\dagger(x_i)\mathcal N^{0,\mathcal{T}}_i\mathcal H_i(x_i)]^{-1} x_i$, local controllers $u_i = \mathcal{U}^{0,\mathcal{T}}_i \mathcal{H}_i (x_i)[\Theta_i^\dagger(x_i)\mathcal N^{0,\mathcal{T}}_i\mathcal H_i(x_i)]^{-1} x_i$, and matrices $\bar{\mathds Z}^{11}_i$, $\bar{\mathds Z}^{12}_i$, $\bar{\mathds Z}^{21}_i$, $\bar{\mathds Z}^{22}_i$, where $\bar{\mathds Z}^{22}_i\preceq 0$ is symmetric\footnotemark[2] 
			\STATE Using the constructed $\mathcal{S}_i(x_i)$, employ \textsf{SOSTOOLS} and satisfy conditions~\eqref{SOS0} and\eqref{SOS1} while designing $\eta_i$ and $\mu_i$
			\ENDFOR
			\STATE Compute supply rate matrices $\mathds Z^{11}_i = \bar{\mathds Z}^{11}_i, \mathds Z^{12}_i = \bar{\mathds Z}^{12}_i[\Theta_i^\dagger(x_i)\mathcal N^{0,\mathcal{T}}_i\mathcal H_i(x_i)]^{-1}$, $\bar{\mathds Z}^{21}_i = \bar{\mathds Z}^{{12}^\top}_i\!\!,$ and $\mathds Z^{22}_i = [\Theta_i^\dagger(x_i)\mathcal N^{0,\mathcal{T}}_i\mathcal H_i(x_i)]^{-1}\bar{\mathds Z}^{22}_i[\Theta_i^\dagger(x_i)\mathcal N^{0,\mathcal{T}}_i\mathcal H_i(x_i)]^{-1}$
			\STATE Compute level sets $\eta=\sum_{i=1}^N \eta_i$ and $\mu=\sum_{i=1}^N \mu_i$
			\IF{compositionality conditions~\eqref{dissi}-\eqref{LMI} are met,}
			\STATE $\mathcal{B}(x)$ = $\sum_{i=1}^N\mathcal{S}_i(x_i)$ with $\mathcal{S}_i(x_i) = x_i^\top {[\Theta_i^\dagger(x_i)\mathcal N^{0,\mathcal{T}}_i\mathcal H_i(x_i)]^{-1}} x_i$ is a CBC for the interconnected network and $u=[u_1; \cdots; u_N]$ with $u_i = \mathcal U^{0,\mathcal{T}}_i \mathcal H_i(x_i)[\Theta_i^\dagger(x_i)\mathcal N^{0,\mathcal{T}}_i\mathcal H_i(x_i)]^{-1}x_i$, $i\in\{1,\dots, N\}$, is its safety controller 
			\ELSE
			\STATE 
			Repeat Step~\ref{Step1} using the enlarged sample set: recompute $
			\mathcal{H}_i(x_i),\quad \bar{\mathds{Z}}_i^{11},\;\bar{\mathds{Z}}_i^{12},\;\bar{\mathds{Z}}_i^{21},\;\bar{\mathds{Z}}_i^{22},
			$
			and verify that the compositionality conditions \eqref{dissi}–\eqref{LMI} are satisfied.
			\ENDIF
			\ENSURE CBC $\mathcal{B}(x)$ = $\sum_{i=1}^N\mathcal{S}_i(x_i)$ with $\mathcal{S}_i(x_i) = x_i^\top [\Theta_i^\dagger(x_i)\mathcal N^{0,\mathcal{T}}_i\mathcal H_i(x_i)]^{-1} x_i$, and its safety controller $u=[u_1; \cdots; u_N]$ with $u_i = \mathcal U^{0,\mathcal{T}}_i \mathcal H_i(x_i)[\Theta_i^\dagger(x_i)\mathcal N^{0,\mathcal{T}}_i\mathcal H_i(x_i)]^{-1}x_i$, for any $i\in\{1,\dots, N\}$
	\end{algorithmic}
\end{algorithm}

\footnotetext[1]{To fulfill condition $\mathcal{N}_i^{0,\mathcal{T}}\mathcal{H}_i(x_i) =\Theta_i(x_i) P_i^{-1}$, we consider $\Theta^\dagger_i(x_i)\mathcal{N}_i^{0,\mathcal{T}}\mathcal{H}_i(x_i) = \mathbf{S}$, with $\mathbf{S}$ being a symmetric positive-definite matrix. Since $\mathbf{S} = P^{-1}_i$ according to \eqref{matrixP}, we compute $P_i$ as the inverse of $\mathbf{S}$ (\emph{i.e.,} $P_i = \mathbf{S}^{-1}$), automatically ensuring that $P_i$ is also positive definite.}
\footnotetext[2]{It is clear that  $\bar{\mathds Z}^{22}_i\preceq 0$\ implies ${\mathds Z}^{22}_i\preceq 0$ since ${\mathds Z}^{22}_i = P_i\bar{\mathds Z}^{22}_iP_i$ and $P_i$ is symmetric, \emph{i.e.,} ${\mathds Z}^{22}_i = P^\top_i \bar{\mathds Z}^{22}_iP_i$. Additionally, imposing symmetry on $\bar{\mathds Z}^{22}_i$ ensures that ${\mathds Z}^{22}_i$ is also symmetric. This holds true since ${\mathds Z}^{22^\top}_i = (P_i\bar{\mathds Z}^{22}_iP_i)^\top = P^\top_i \bar{\mathds Z}^{22^\top}_i P^\top_i = P_i \bar{\mathds Z}^{22}_i P_i = {\mathds Z}^{22}_i$, when $ \bar{\mathds Z}^{22^\top}_i =  \bar{\mathds Z}^{22}_i$.}

\begin{remark}
Condition \eqref{LMI} represents a well-defined linear matrix inequality (LMI) delineated in \citep{arcak2016networks}, initially employed as a compositional criterion in dissipativity theory to demonstrate the stability of interconnected networks based on their subsystems. We utilize this condition within our framework to illustrate how a CBC can be designed based on CSCs of individual subsystems, derived from data. As discussed by \cite{arcak2016networks}, this condition can leverage the structure of the interconnection topology and be satisfied irrespective of subsystems' number for specific types of interconnection structures, such as skew symmetry, i.e., $\mathds{M}^{\top} = -\mathds{M} $. This offers that skew-symmetric interconnection structures are not a restrictive assumption but rather a beneficial property within the compositional framework.
\end{remark}

\begin{remark}
An advantage of our compositional framework for designing safety controllers within interconnected networks is that it does not require the enforcement of condition \(\mu_i > \eta_i\) for each subsystem. Instead, it is adequate to impose the condition \( \mu > \eta \) (cf. condition~\eqref{dissi}) for the entire network, in conjunction with condition~\eqref{LMI} for constructing CBC and its safety controller. This relaxation effectively diminishes the conservatism of the constraints outlined in Theorem~3.4, thereby enhancing the likelihood of finding appropriate CSC and local controllers for each subsystem. In case of a homogeneous interconnected network, where all subsystems share identical characteristics, satisfying the condition \( \mu_i > \eta_i \) for just one subsystem ensures that the condition \( \mu > \eta \) is also fulfilled (cf. condition~\eqref{dissi}). However, in general, the required compositional condition $\sum_{i=1}^N \eta_i < \sum_{i=1}^N \mu_i$ allows some subsystems to violate $\mu_i > \eta_i$, highlighting the flexibility of the compositional reasoning in the context of barrier certificate construction.
\end{remark}

We present Algorithm~\ref{Alg:1} to delineate the essential steps for designing a CBC and its corresponding safety controller across an \emph{unknown} interconnected network.\vspace{-0.2cm}

\section{Case study: A set of benchmarks}\label{case-study}
We demonstrate the efficacy of our proposed findings by applying them to a set of benchmarks, including interconnected networks of Lorenz, Chen~\citep{lopez2019synchronization},  spacecraft~\citep{khalil2002control}, and Duffing oscillator~\citep{WANG2014162} systems, connected through various interconnection topologies (see Fig. \ref{topology2}). A concise overview of these case studies is provided in Table~\ref{tab:system-configurations}. Notably, Lorenz-type systems (\emph{i.e.,} Lorenz, Chen) are \emph{chaotic} and widely used in real-world applications due to their ability to model complex, chaotic behaviors. These applications include secure communications for encryption using chaotic signals~\citep{Wang2009}, weather prediction models to simulate atmospheric dynamics~\citep{DeterministicNonperiodicFlow}, robotics and autonomous systems adapting to unpredictable environments~\citep{sprott2010elegant}, and neuroscience for modeling chaotic brain activity and understanding neural dynamics and disorders such as epilepsy~\citep{strogatz2018nonlinear}.

\begin{table*}[t!]
\makeatletter
\long\def\@makecaption#1#2{%
	\vskip\abovecaptionskip
	\noindent \textbf{#1.} #2\par
	\vskip\belowcaptionskip}
\makeatother
\caption{An overview of interconnected network configurations and parameters: $N$ indicates the number of subsystems, \(\mathcal{T}\) represents the horizon of data samples, $n_i$ specifies the number of states of each subsystem, \(\bar{\phi}_i\) denotes the noise bound coefficient as outlined in \eqref{noise-bound-2}, \(\eta_i\) and \(\mu_i\) specify the initial and unsafe level sets, $\lambda_i$ is decay rate, \textsf{RT} denotes running time~\textsf{seconds}\protect\footnotemark[3], and \textsf{MU} indicates memory usage~\textsf{Mbit}, required for each subsystem. As can be seen, under strongly connected topologies (\emph{e.g.,} fully interconnected), the number of subsystems is smaller compared to other topologies, since each subsystem influences all others, making the compositionality condition~\eqref{LMI} more constraining.
\label{tab:system-configurations}}
\centering
\begin{tabular}{@{}llcccccccccc@{}}
		\toprule
		\textbf{Type of System} & \textbf{Topology} & $N$ & $\mathcal{T}$ & $n_i$ & $\bar{\phi}_i$ & $\eta_i$ & $\mu_i$ &  $\lambda_i$ & \textsf{RT} & \textsf{MU} \\ 
		\midrule
		\midrule
  	\multirow{2}{*}{Lorenz}      &Fully connected & 1000& 15 & 3& 0.03&  98.21  &    100.61 & 0.99&16.08 & 35.16   \\            
   & Ring & 1500 & 13 &3& 0.12 &  123.17  & 126.39 & 0.99& 11.96&  29.05  \\ 
    \hline
		\multirow{2}{*}{Spacecraft} & Binary& 1023\protect\footnotemark[4]& 15 &3 & 0.75 &  87.17  &  89.27 & 0.99& 14.49 & 35.1 \\
		&Star& 2000& 17& 3& 0.75&  128.01 & 132.68 & 0.99& 19.46 & 41.92   \\
		\hline
		Chen     & Line              & 1000 & 12  & 3& 0.27 & 35.63 &   36.16 & 0.99& 9.05 & 26.2 \\
		\hline
		Duffing oscillator             & Binary                & 1023 & 18 & 2& 0.08&  406.61&  412.52 & 0.99& 3.27 &  15.24  \\
		\bottomrule
	\end{tabular}
\end{table*}
The primary objective in all benchmarks is to design a CBC and its safety controller for the interconnected network with an \emph{unknown} mathematical model, ensuring that the system's states remain within a safe region over an infinite time horizon. To achieve this, under Algorithm~\ref{Alg:1}, we collect input-state trajectories as outlined in~\eqref{New} and meet conditions~\eqref{SOS6} by constructing CSCs and local controllers for each subsystem. Following the compositionality findings of Theorem~\ref{dissipativity}, we then compositionally construct a CBC and its safety controller for the interconnected networks, thereby ensuring the network safety. It is noteworthy that all simulations were performed on a $\mathsf{MacBook}$ with an $\mathsf{M2 \, chip}$ and $32$ GB of memory.

The following subsections offer comprehensive descriptions of the benchmarks presented in Table~\ref{tab:system-configurations}. This encompasses the regions of interest, the dictionaries of monomials up to a specified degree, the network topology, and the subsystem matrices \(A_i\), \(B_i\) and \(D_i\) in addition to the overall interconnected network matrices \(A(x)\) and \(B\) with the coupling block matrix $\mathds{M}$. It is crucial to highlight that the matrices \(A_i\) and \(B_i\) are considered unknowns in all case studies. Additionally, details regarding CSC and their respective local controllers \(u_i\) for subsystems, as well as CBC and its safety controller \(u\) for the interconnected network, along with the network parameters (\emph{i.e.,} $\eta$, $\mu$, and $\lambda$), are reported in the Appendix. The supply rate matrices ${\mathds Z}^{11}_i,{\mathds Z}^{12}_i,{\mathds Z}^{21}_i,$ and ${\mathds Z}^{22}_i$ are not reported  for brevity.
\begin{figure}[t!]
	\centering
	\begin{subfigure}[b]{0.33\linewidth}\label{a2}
		\centering
		\resizebox{0.8\linewidth}{!}{%
			\begin{tikzpicture}
				\tikzset{vertex/.style = {
						shape=circle,
						draw,
						ball color=purple!5,
						minimum size=1.5em
				}}
				\tikzset{edge/.style = {draw, thick}}
				\node[vertex] (a) at (0:2) {{$\Xi_1$}};
				\node[vertex] (b) at (60:2) {{$\Xi_2$}};
				\node[vertex] (c) at (120:2) {{$\Xi_3$}};
				\node[vertex] (d) at (180:2) {{$\Xi_4$}};
				\node[vertex] (e) at (240:2) {{$\Xi_5$}};
				\node[vertex] (f) at (300:2) {{$\Xi_{N}$}};
				\draw[edge] (a) -- (b);
				\draw[edge] (b) -- (c);
				\draw[edge] (c) -- (d);
				\draw[edge] (d) -- (e);
				\draw[edge, dash dot] (e) -- (f);
				\draw[edge] (f) -- (a);
				\draw[edge] (a) -- (c);
				\draw[edge] (a) -- (d);
				\draw[edge] (a) -- (e);
				\draw[edge] (b) -- (d);
				\draw[edge] (b) -- (e);
				\draw[edge] (b) -- (f);
				\draw[edge] (c) -- (e);
				\draw[edge] (c) -- (f);
				\draw[edge] (d) -- (f);
				\draw[edge ] (e) -- (a);
			\end{tikzpicture}
		}
		\caption{Fully connected}
		\label{fully}
	\end{subfigure}
	\hspace{1cm}\begin{subfigure}[b]{0.43\linewidth}
		\centering
		\resizebox{0.8\linewidth}{!}{%
			\begin{tikzpicture}[level/.style={sibling distance = 4cm/#1,
					level distance = 1.5cm}]
				\tikzset{vertex/.style = {shape=circle,,fill=green!40,draw,ball color= green!20, minimum size=1.5em}}
				\tikzset{edge/.style = {->, draw, thick}}  
				\tikzset{dashdot/.style = {draw, thick, dash dot, ->}}
				\node [vertex] (a) {{$\Xi_1$}}
				child {node [vertex] (b) {{$\Xi_2$}}[->, thick]
					child {node [vertex] (d) {{$\Xi_4$}}[dashdot]
					}
					child {node [vertex] (e) {{$\Xi_5$}}[dashdot]
					}
				}
				child {node [vertex] (c) {$\Xi_3$}[->, thick]
					child {node [vertex] (f) {{$\Xi_6$}}[dashdot]
					}
					child {node [vertex] (g) {{$\Xi_7$}}[dashdot]
					}
				};
			\end{tikzpicture}
		}
		\caption{Binary}
		\label{binary}
	\end{subfigure}\\
	\begin{subfigure}[b]{0.33\linewidth}
		\centering
		\resizebox{0.8\linewidth}{!}{%
			\begin{tikzpicture}
				\tikzset{vertex/.style = {shape=circle,draw, fill=pink!70, ball color= pink!20, minimum size=1.5em}}
				\tikzset{edge/.style = {draw, thick, ->}}
				\node[vertex] (center) at (0,0) {$\Xi_1$};
				\node[vertex] (1) at (0:2) {$\Xi_2$};
				\node[vertex] (2) at (72:2) {$\Xi_3$};
				\node[vertex] (3) at (144:2) {$\Xi_4$};
				\node[vertex] (4) at (216:2) {$\Xi_5$};
				\node[vertex] (5) at (288:2) {$\Xi_N$};
				\draw[edge] (center) -- (1);
				\draw[edge] (center) -- (2);
				\draw[edge] (center) -- (3);
				\draw[edge] (center) -- (4);
				\draw[edge, dash dot] (center) -- (5);
			\end{tikzpicture}
		}
		\caption{Star}
		\label{star}
	\end{subfigure}
	\hspace{1cm}\begin{subfigure}[b]{0.33\linewidth}
		\centering
		\resizebox{0.8\linewidth}{!}{%
			\begin{tikzpicture}
				\tikzset{vertex/.style = {shape=circle,draw, fill=yellow!50,ball color= yellow!30, minimum size=1.5em}}
				\tikzset{edge/.style = {draw, thick, ->}}
				\node[vertex] (1) at ({360/8 * (1 - 1)}:2cm) {$\Xi_1$};
				\node[vertex] (2) at ({360/8 * (2 - 1)}:2cm) {$\Xi_2$};
				\node[vertex] (3) at ({360/8 * (3 - 1)}:2cm) {$\Xi_3$};
				\node[vertex] (4) at ({360/8 * (4 - 1)}:2cm) {$\Xi_4$};
				\node[vertex] (5) at ({360/8 * (5 - 1)}:2cm) {$\Xi_5$};
				\node[vertex] (6) at ({360/8 * (6 - 1)}:2cm) {$\Xi_6$};
				\node[vertex] (7) at ({360/8 * (7 - 1)}:2cm) {$\Xi_7$};
				\node[vertex] (8) at ({360/8 * (8 - 1)}:2cm) {$\Xi_N$};
				\draw[edge] (1) -- (2);
				\draw[edge] (2) -- (3);
				\draw[edge] (3) -- (4);
				\draw[edge] (4) -- (5);
				\draw[edge] (5) -- (6);
				\draw[edge] (6) -- (7);
				\draw[edge, dash dot] (7) -- (8);
				\draw[edge] (8) -- (1); 
			\end{tikzpicture}
		}
		\caption{Ring}
		\label{ring}
	\end{subfigure}\\
	\vspace{0.2cm}\begin{subfigure}[b]{0.33\linewidth}
		\centering
		\resizebox{\linewidth}{!}{%
			\begin{tikzpicture}[node distance=1.8cm]
				\tikzset{vertex/.style = {shape=circle, draw, align=center, minimum size=1.5em,ball color= blue!20, fill=blue!30}}
				\tikzset{edge/.style = {draw, thick, ->}}
				\node[vertex] (1) {$\Xi_1$};
				\node[vertex] (2) [right of=1] {$\Xi_2$};
				\node[vertex] (3) [right of=2] {$\Xi_3$};
				\node[vertex] (4) [right of=3] {$\Xi_N$};
				\draw[edge] (1) -- (2);
				\draw[edge] (2) -- (3);
				\draw[edge, dash dot] (3) -- (4);
			\end{tikzpicture}
		}
		\caption{Line}
		\label{line}
	\end{subfigure}
	\caption{Various interconnection topologies utilized in Table \ref{tab:system-configurations} for different benchmarks.\label{topology2}}
\end{figure}
\subsection{Lorenz network with fully-interconnected topology}
We consider the Lorenz network with fully interconnected topology, in which each subsystem has $3$ state variables $x_i=\left[x_{i_1} ; x_{i_2}; x_{i_3}\right]$, for all $i \in \{1,\ldots,1000\}$, whose dynamics are described as
	\begin{align}\notag
			& \dot{x}_{i_1}=10 \,x_{i_2}-10\, x_{i_1} - 10^{-5} \!\!\!\! \sum_{\substack{j=1, j \neq i}}^{1000} x_{j_1}  \!+\! u_{i_1},\\\notag
			& \dot {x}_{i_2}=28\, x_{i_1}\!-\! x_{i_2} \!- \! x_{i_1} x_{i_3}\! -\!  10^{-5}\!\!\!\! \sum_{\substack{j=1, j \neq i}}^{1000} x_{j_2} \!+\! u_{i_2}, \\\label{lorenz-fully}
			& \dot{x}_{i_3}=x_{i_1} x_{i_2}-\frac{8}{3}\, x_{i_3}- 10^{-5}\!\!\!\! \sum_{\substack{j=1, j \neq i}}^{1000} x_{j_3}  \! + \! u_{i_3}.
	\end{align}
	One can rewrite the dynamics in~\eqref{lorenz-fully} in the form of \eqref{sys2} with actual $\mathcal{R}_i(x_i) = [x_{i_1}; x_{i_2}; x_{i_3}; x_{i_1}x_{i_3}; x_{i_1}x_{i_2}]$, and
	\begin{align*}
	A_i &= \begin{bmatrix} -10&& 10 && 0 && 0 && 0  \\28 && -1 && 0 && -1 && 0  \\ 0 && 0 && -\frac{8}{3} && 0 && 1 \end{bmatrix}\!\!,~
	B_i = \mathbf{I}_{3},~
	D_i =  -10^{-5} \,\mathbf{I}_{3}.
	\end{align*}
   It is worth re-emphasizing that the matrices $A_i$ and $B_i$ are assumed to be unknown. While the exact form of $\mathcal{R}_i(x_i)$ is also unknown, we specify a dictionary of monomials up to degree 2 as $\mathcal{R}_i(x_i)=[x_{i_1}; x_{i_2}; x_{i_3}; x_{i_1}x_{i_3}; x_{i_1}x_{i_2}; x_{i_2}x_{i_3}; x_{i_1}^2; x_{i_2}^2; x_{i_3}^2]$.  Under the fully interconnected topology as
	\begin{align*}
    \mathds{M} &= \begin{bmatrix}
		\mathbf{0}_{3\times3} && \mathbf{I}_{3} &&\mathbf{I}_{3} &&\ldots && \mathbf{I}_{3}\\
	    \mathbf{I}_{3} &&	\mathbf{0}_{3\times3} && \mathbf{I}_{3} &&\ldots &&\mathbf{I}_{3}\\
		\vdots&&\quad &&\ddots&&\quad &&\vdots \\
		\mathbf{I}_{3}&&\ldots &&\mathbf{I}_{3}&&	\mathbf{0}_{3\times3} &&\mathbf{I}_{3} \\
		\mathbf{I}_{3} &&\ldots &&\mathbf{I}_{3} && \mathbf{I}_{3} &&	\mathbf{0}_{3\times3}
	\end{bmatrix}_{3000\times 3000}\!\!\!\!\!\!\!\!\!\!\!\!\!\!\!\!\!\!\!\!\!\!\!\!\!,
	\end{align*}	
	the matrices of the interconnected network can be structured as 
	\begin{align*}
		 A(x)~\text{as in ~\eqref{A(x)-lorenz-fully}}, ~~B= \mathsf{blkdiag}(B_1,\ldots,B_{1000}).
	\end{align*}
    The regions of interest are also given by the state set $X_i = [-20,20]^3$, the initial set $X_{0_i} = [-3,3]^3$, and the unsafe set $X_{a_i} = [-20,-4]\times[-20,-15]\times[4,20] \cup [5,20]\times[11,20]\times[4,20] \cup [5,20]\times[11,20]\times[-20,-4]$ for all $i \in \{1,\ldots,1000\}$.
	\begin{figure*}[!t] 
			\rule{\textwidth}{0.1pt}
\begin{align}\label{A(x)-lorenz-fully}
	A(x) &= \begin{bmatrix}
		A_1 \Theta_1(x_1) & D_{1} & D_{1} &  \ldots  & D_{1} \\
		D_{2}& A_2 \Theta_2(x_2)  & D_{2} &  \ldots  &  D_{2} \\
		\vdots \!\! & \!\!\quad \!\! & \!\!\ddots \!\! & \!\!\quad \!\! & \!\!\vdots \\
		D_{999}  & \ldots  & D_{999}  & A_{999} \Theta_{999}(x_{999})  & D_{999} \\
		D_{1000}  & \ldots & D_{1000}  & D_{1000} & A_{1000}\Theta_{1000}(x_{1000})
	\end{bmatrix}
\end{align}
\end{figure*}
\begin{figure*}[!t] 
\begin{align}\label{A(x)-lorenz-ring}
			A(x) &\!=\! \begin{bmatrix}
				A_1 \Theta_1(x_1)  & \mathbf{0}_{3 \times 3}  & \mathbf{0}_{3 \times 3} & \cdots  & D_{1}\\
				D_{2}  & A_2 \Theta_2(x_2)  & \mathbf{0}_{3 \times 3}  & \cdots  & \mathbf{0}_{3 \times 3} \\
				\vdots \!\! & \!\! \!\! & \!\!\ddots \!\! & \!\! \!\! & \!\!\vdots \\
				\mathbf{0}_{3 \times 3}  & \cdots  &  D_{1499} & A_{1499} \Theta_{1499}(x_{1499}) & \mathbf{0}_{3 \times 3} \\
				\mathbf{0}_{3 \times 3}  & \cdots  & \mathbf{0}_{3 \times 3} & D_{1500}  & A_{1500} \Theta_{1500}(x_{1500})
			\end{bmatrix}
	\end{align}
	\rule{\textwidth}{0.1pt}
\end{figure*}
\footnotetext[3]{The specified memory usage and running time are for solving the SOS problem of Algorithm~\ref{Alg:1}.}
\footnotetext[4]{The number of subsystems, denoted as \( \boldmath{N} \), in a binary topology is given by \( \boldmath{N} = 2^\ell - 1 \), where \( \ell \) represents the number of layers in the binary topology.}
\subsection{Lorenz network with ring interconnection topology}
We analyze the Lorenz network with a ring topology, where  its dynamics for \(i = \{2, \dots, 1500\}\) are described by
\begin{align}\notag
	& \dot{x}_{i_1}=10\, x_{i_2}-10\, x_{i_1} -0.08\,{x}_{(i-1)_1}+ u_{i_1},\\\notag
	& \dot {x}_{i_2}=28\, x_{i_1}-x_{i_2}-x_{i_1} x_{i_3}- 0.08\,{x}_{(i-1)_2} + u_{i_2}, \\\label{lorenz-ring}
	& \dot{x}_{i_3}=x_{i_1} x_{i_2}-\frac{8}{3}\, x_{i_3}- 0.08\,{x}_{(i-1)_3}+ u_{i_3},
	\end{align}
	with the first subsystem being affected by the last one. One can rewrite the dynamics in~\eqref{lorenz-ring} in the form of \eqref{sys2} with actual $\mathcal{R}_i(x_i) = [x_{i_1}; x_{i_2}; x_{i_3}; x_{i_1}x_{i_3}; x_{i_1}x_{i_2}]$ and
	\begin{align*}
	A_i &= \begin{bmatrix} -10& 10 & 0 & 0 & 0  \\28 & -1 & 0 & -1& 0 \\ 0 & 0 & -\frac{8}{3} & 0 & 1 \end{bmatrix}\!\!,\,
	B_i = \mathbf{I}_3,\, 	D_i = -0.08\,\mathbf{I}_3,\\
		\end{align*}
		where both matrices \(A_i\) and \(B_i\) are treated as unknown. Likewise, while the exact form of \(\mathcal{R}_i(x_i)\) is unknown, we construct a dictionary of monomials up to degree $2$ as \(\mathcal{R}_i(x_i) = [x_{i_1}; x_{i_2}; x_{i_3}; x_{i_1}x_{i_3}; x_{i_1}x_{i_2}; x_{i_2}x_{i_3}; x_{i_1}^2; x_{i_2}^2; x_{i_3}^2]\). Under the \emph{ring} topology as
		\begin{align*}
    	\mathds{M} &= \begin{bmatrix}
		 \mathbf{0}_{3 \times 3} && \mathbf{0}_{3 \times 3} && \mathbf{0}_{3 \times 3} && \cdots  && \mathbf{I}_{3}\\
	     \mathbf{I}_{3}  &&  \mathbf{0}_{3 \times 3} && \mathbf{0}_{3 \times 3}  && \cdots && \mathbf{0}_{3 \times 3} \\
			\vdots&&\quad &&\ddots&&\quad &&\vdots \\
		\mathbf{0}_{3 \times 3}  && \cdots  &&  \mathbf{I}_{3}   &&  \mathbf{0}_{3 \times 3} && \mathbf{0}_{3 \times 3} \\
		\mathbf{0}_{3 \times 3}  && \cdots & & \mathbf{0}_{3 \times 3} &&  \mathbf{I}_{3}   &&  \mathbf{0}_{3 \times 3}
	\end{bmatrix}_{4500\times 4500}\!\!\!\!\!\!\!\!\!\!\!\!\!\!\!\!\!\!\!\!\!\!\!\!\!,
	\end{align*}
the matrices of the interconnected network can be written as 	
\begin{align*}
A(x)~\text{as in ~\eqref{A(x)-lorenz-ring}}, ~~B= \mathsf{blkdiag}(B_1,\ldots,B_{1500}).
\end{align*}
The regions of interest are specified by the state set $X_i = [-20, 20]^3$, the initial set $X_{0_i} = [-3,3]^3$, and the unsafe set $X_{a_i} = [-20,-10] \times[-20,-5] \times [5, 20] \cup [3.5,20] \times[15,20] \times [5, 20] \cup [3.5,20] \times[15, 20] \times [-20, -5]$  for all $i \in \{1,\ldots,1500\}.$

\begin{figure*}[!t]
	\centering
	\centering
	\begin{subfigure}[t]{0.24\textwidth}
		\includegraphics[width=\linewidth]{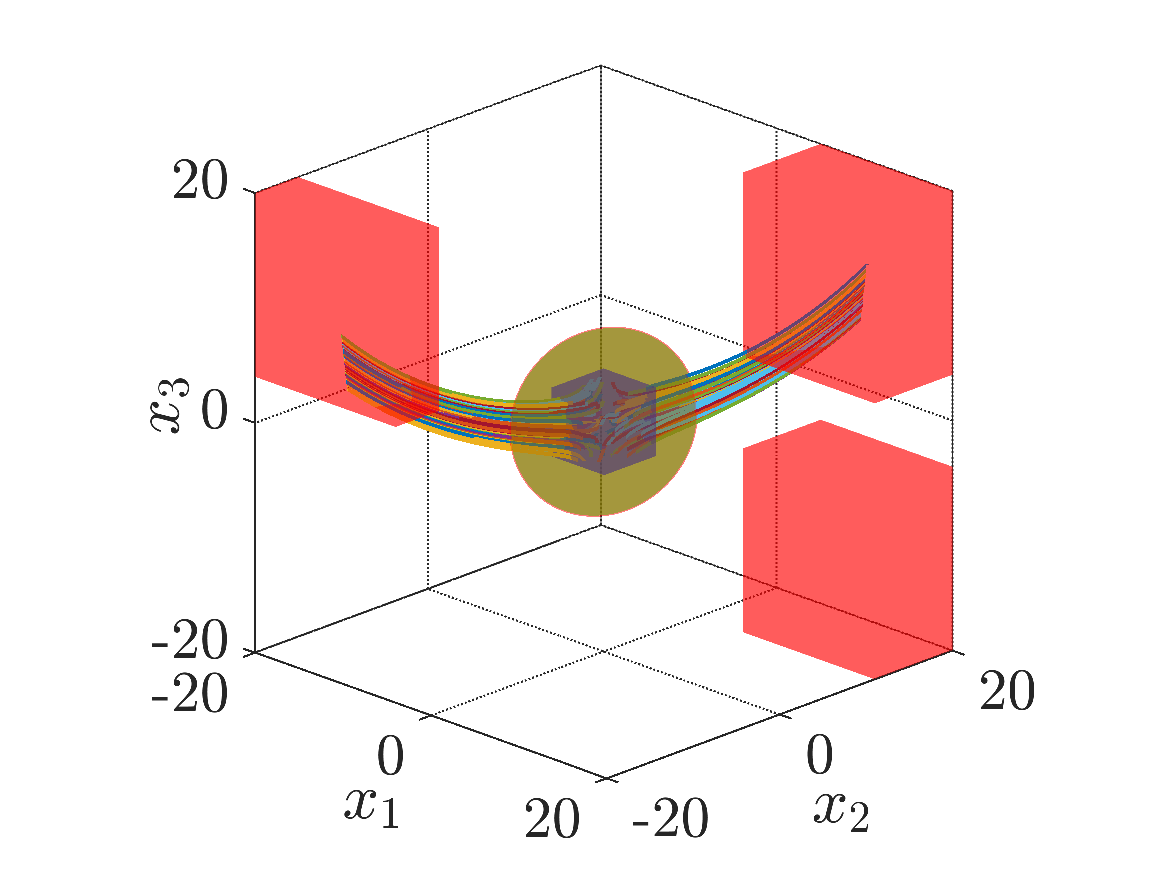}
		\caption{Open-loop trajectories}
	\end{subfigure}
	\hspace{-0.4cm}
	\begin{subfigure}[t]{0.24\textwidth}
		\includegraphics[width=\linewidth]{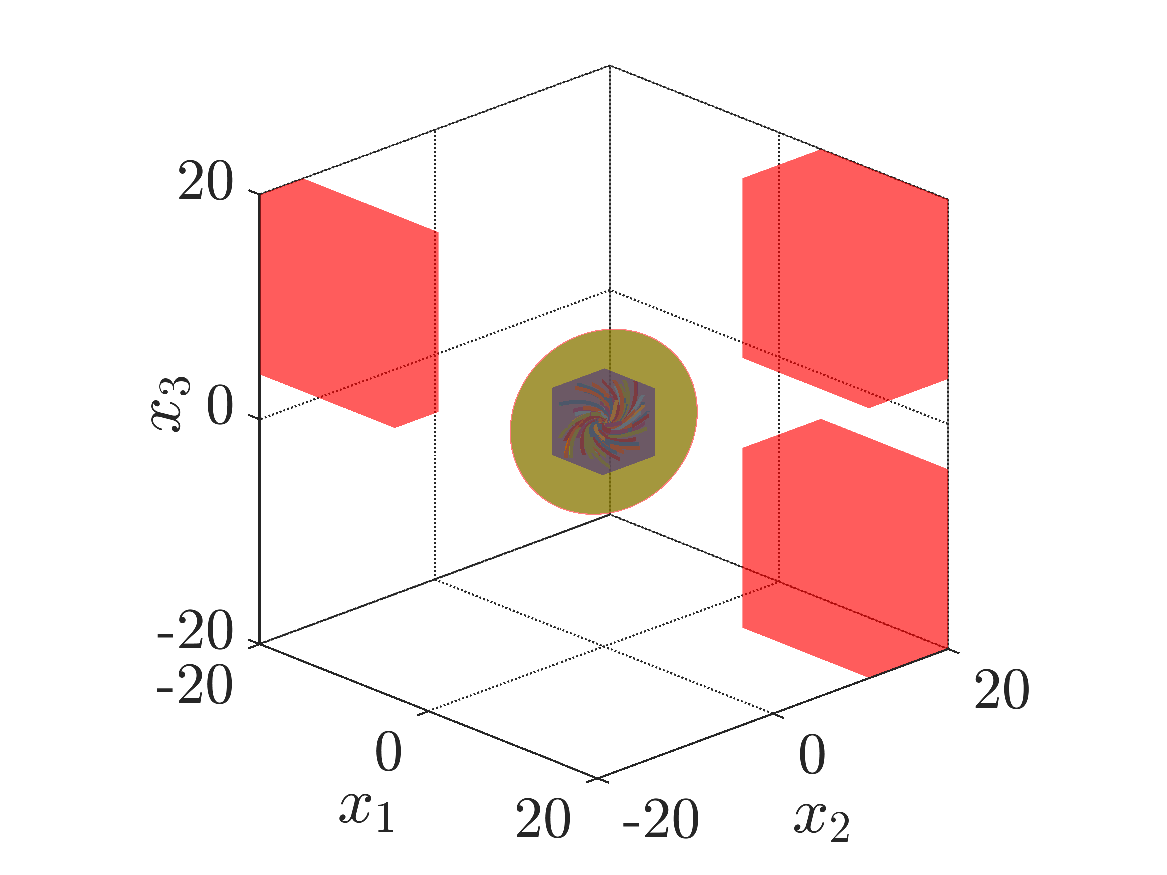}
		\caption{Closed-loop trajectories}
	\end{subfigure}
	\begin{subfigure}[t]{0.24\textwidth}
		\includegraphics[width=\linewidth]{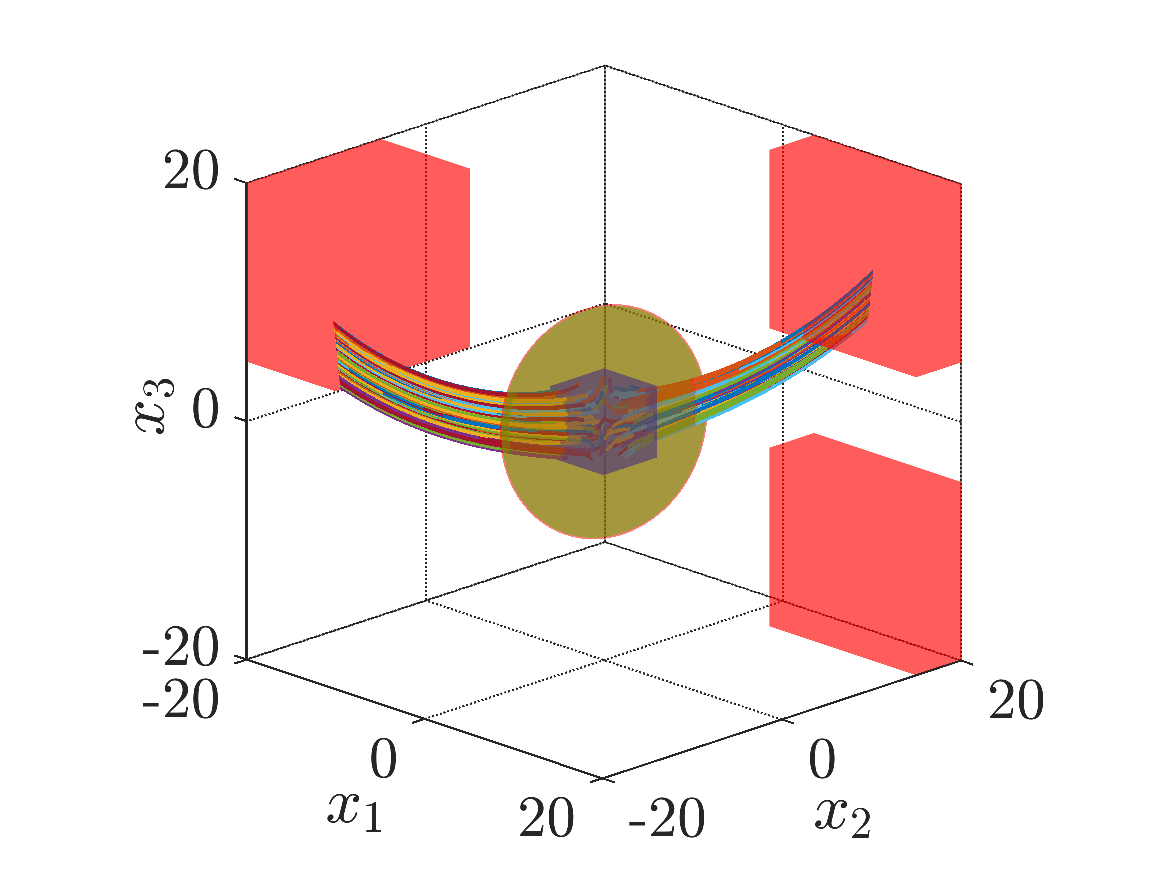}
		\caption{Open-loop trajectories}
	\end{subfigure}
	\hspace{-0.4cm}
	\begin{subfigure}[t]{0.24\textwidth}
		\includegraphics[width=\linewidth]{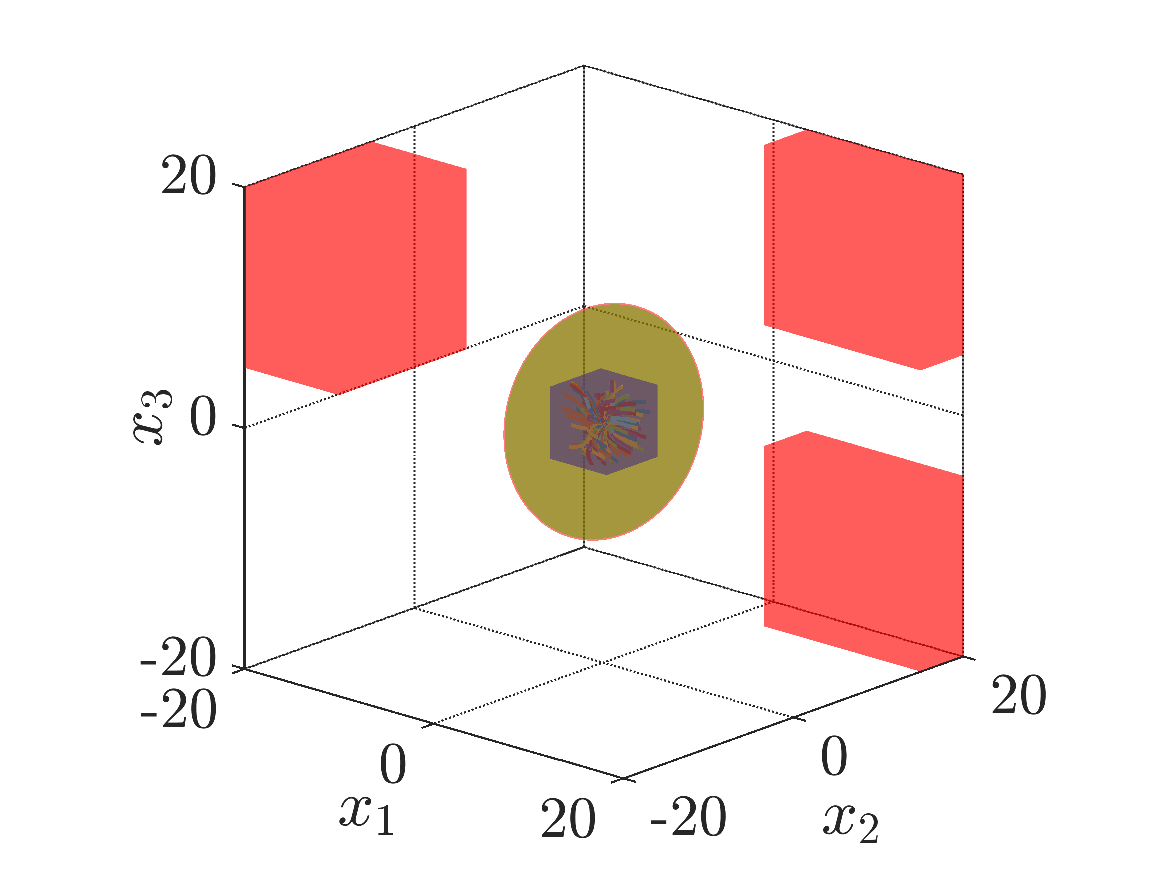}
		\caption{Closed-loop trajectories}
	\end{subfigure}
	\caption{Trajectories of $120$ representative
			Lorenz subsystems in the network of $1000$ and $1500$ components with fully-connected \textbf{(a),(b)} and ring \textbf{(c),(d)} interconnection topologies, respectively. Green \protect\greensquare\ and pink \protect\pinksquare\ surfaces represent initial and unsafe level sets $\mathcal B (x) = \eta$ and $\mathcal B (x) = \mu$, respectively (due to the close proximity of the level sets, the unsafe level set $\mathcal{B}(x)=\mu$, indicated in pink, is not clearly visible). Initial and unsafe regions are depicted by blue \protect\bluesquare\ and red \protect\redsquare\ boxes. Figures \textbf{(a),(c)}  indicate the open-loop trajectories of the Lorenz networks with respective topologies. As can be seen, due to the \emph{chaotic} nature of the Lorenz networks, trajectories of representative subsystems without a safety controller immediately violate safety specifications.}
	\label{fig:4}
\end{figure*}

Trajectories of representative Lorenz subsystems with fully-connected and ring interconnection topologies are depicted in Figure \ref{fig:4}.

\subsection{Spacecraft network with binary interconnection topology}
We examine a spacecraft  network with the binary topology, in which each spacecraft has $3$ states $x_i=\left[x_{i_1}; x_{i_2}; x_{i_3}\right]$, which are the angular velocities $\omega_ {i_1}$ to $\omega_{i_3}$ along the principal axes. The spacecraft's dynamics are characterized by
\begin{align}\notag
	\dot{x}_{i_1} &= \frac{J_{i_2} - J_{i_3}}{J_{i_1}}\, x_{i_2}\, x_{i_3} 
	+ \frac{1}{J_{i_1}}\, u_{i_1} 
	\\\notag&~~~+ \frac{0.08}{J_{i_1}}\,\Omega\,(i=2 j \vee i=2 j+1) x_{j_1}, \\\notag 
	\dot{x}_{i_2} &= \frac{J_{i_3} - J_{i_1}}{J_{i_2}}\, x_{i_1}\, x_{i_3} 
	+ \frac{1}{J_{i_2}}\, u_{i_2} 
	\\\notag&~~~+ \frac{0.08}{J_{i_2}}\,\Omega\,(i=2 j \vee i=2 j+1) x_{j_2}, \\\notag
	\dot{x}_{i_3} &= \frac{J_{i_1} - J_{i_2}}{J_{i_3}}\, x_{i_1}\, x_{i_2} 
	+ \frac{1}{J_{i_3}}\, u_{i_3} 
	\\\label{space-line} &~~~+ \frac{0.08}{J_{i_3}}\,\Omega\,(i=2 j \vee i=2 j+1) x_{j_3}, 
	\end{align}
where $\vee$ stands for the logical $\mathsf{OR}$ operation and $\Omega$ is an indicator that we define it as
	\begin{equation*}
		\Omega\,(i=2 j \vee i=2 j+1)= \begin{cases}1, & i=2 j \vee i=2 j+1, \\ 0, & \text { Otherwise}.\end{cases}
\end{equation*}
Here, $u_i=\left[u_{i_1} ; u_{i_2} ; u_{i_3}\right]$ is the torque input, and $J_{i_1}$ to $J_{i_3}$ are the principal moments of inertia. One can rewrite the dynamics in~\eqref{space-line} in the form of \eqref{sys2} with actual $\mathcal{R}_i(x_i) = [x_{i_1},x_{i_2},x_{i_3},   x_{i_2}x_{i_3}; x_{i_3}x_{i_1}; x_{i_1}x_{i_2} ]$ and
	\begin{align*}
	A_i \!&=\! \begin{bmatrix}  0 \quad& 0 \quad& 0 \quad& \frac{J_{i_2} - J_{i_3}}{J_{i_1}}& 0& 0  \\0 \quad& 0 \quad& 0 \quad& 0 \quad& \frac{J_{i_3} - J_{i_1}}{J_{i_2}} \quad& 0 \\0 \quad& 0 \quad& 0 \quad& 0 \quad& 0 \quad& \frac{J_{i_1} - J_{i_2}}{J_{i_3}} \end{bmatrix}\!\!,\\
	B_i \!&= \! \begin{bmatrix} \frac{1}{J_{i_1}}  & 0  & 0\\ 0 &  \frac{1}{J_{i_2}} &0 \\ 0 &0 &\frac{1}{J_{i_3}}    \end{bmatrix}\!\!,
	~~~ D_i \!=\!\!  \begin{bmatrix} \frac{0.08}{J_{i_1}}  & 0  & 0\\ 0 &  \frac{0.08}{J_{i_2}} &0 \\ 0 &0 &\frac{0.08}{J_{i_3}}    \end{bmatrix}\!\!.\\
\end{align*}
It is important to reiterate that the matrices $A_i$ and $B_i$ are considered to be unknown. Although the precise expression of $\mathcal{R}_i(x_i)$ is also not known, the dictionary of monomials up to the degree of $2$ is given by \(\mathcal{R}_i(x_i)= [x_{i_1}; x_{i_2}; x_{i_3}; x_{i_2} x_{i_3}; x_{i_3} x_{i_1}; x_{i_1} x_{i_2};x^2_{i_1}; x^2_{i_2}; x^2_{i_3}]\). Under the \emph{binary} topology as
\begin{align*}
\mathds{M}&=\{\mathbb{m}_{ij}\} = \begin{cases}  \mathbf{0}_{3 \times 3}, & i=j, \\ \mathbf{I}_{3}, & i=2 j \vee i=2 j+1, \\ \mathbf{0}_{3 \times 3}, & \text{Otherwise}, \end{cases}
	\end{align*}
the matrices of the interconnected network can be arranged as 	
\begin{align*}
	A(x)&=\left\{\mathbb{a}_{i j}\right\} = \begin{cases} A_i \Theta_{i}(x_{i}), & i=j, \\ ~~~ D_{i}, & i=2 j \vee i=2 j+1, \\ \mathbf{0}_{3 \times 3}, & \text{Otherwise}, \end{cases}\\
	B &= \mathsf{blkdiag}(B_1,\ldots,B_{1023}).
	\end{align*}
The regions of interest are defined as the state set \(X_i = [-5, 5]^3\), the initial set \(X_{0_i} = [-2,2]^3\), and the unsafe set \(X_{a_i} = [2.5,5] \times [-5,-3] \times [-5,-2.5] \cup [2.5,5] \times [2.5,5] \times [2.5,5] \cup [-5,-3] \times [2.5, 5] \times [2.5,5]\) for all \(i \in \{1,\ldots,1023\}\).
\begin{figure*}
	\centering
		\centering
	\begin{subfigure}[t]{0.28\textwidth}
		\includegraphics[width=\linewidth]{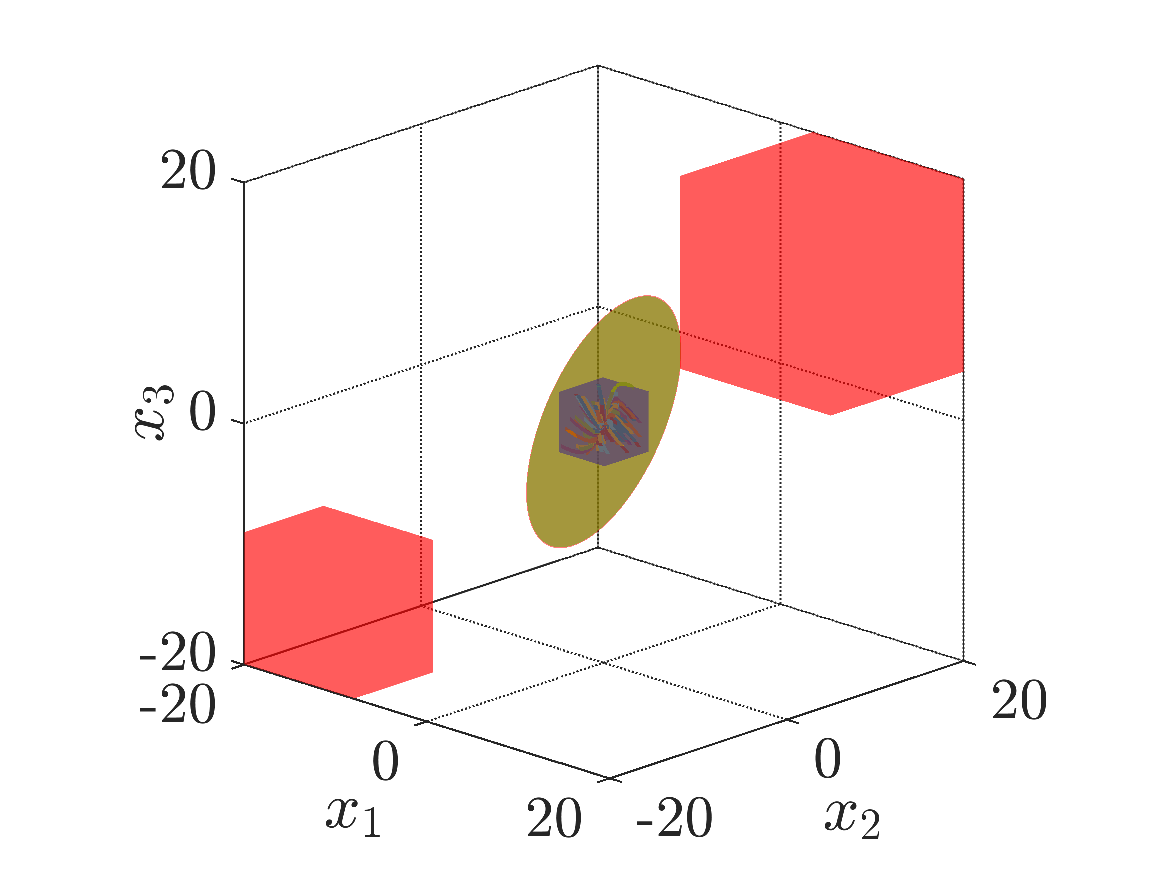}
	\end{subfigure}
	\begin{subfigure}[t]{0.28\textwidth}
	\includegraphics[width=\linewidth]{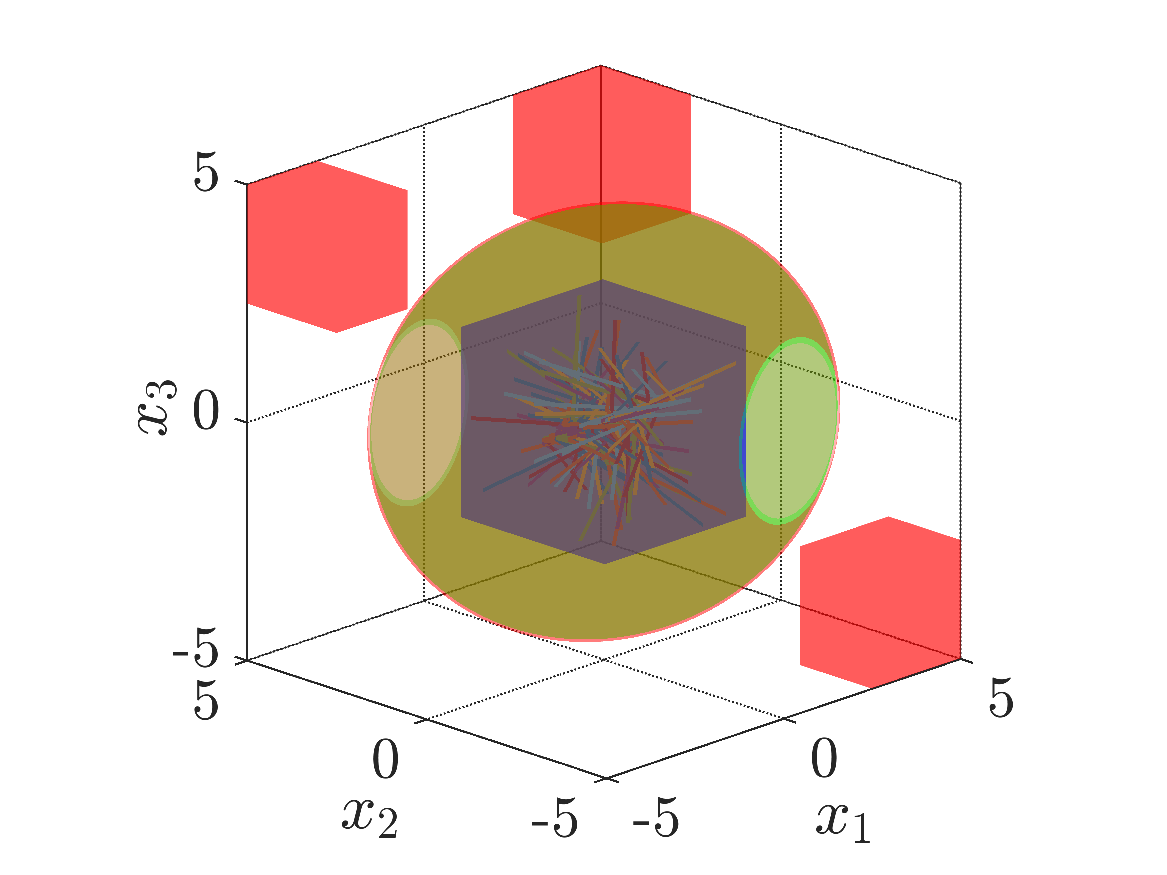}
\end{subfigure}
\hspace{-0.4cm}
\begin{subfigure}[t]{0.28\textwidth}
	\includegraphics[width=\linewidth]{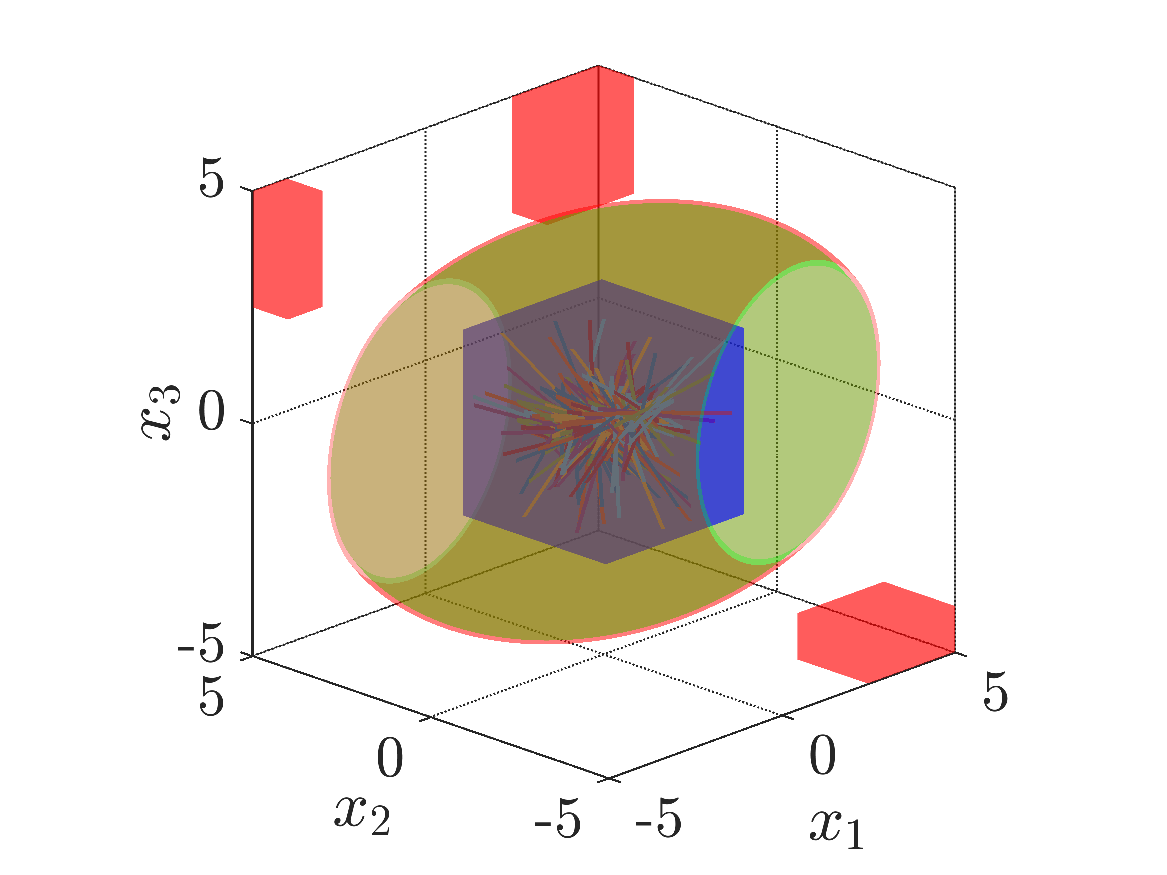}
\end{subfigure}
	\caption{Closed-loop trajectories of $120$ representative
Chen subsystems with line interconnection topology \textbf{(left)} in the network of $1000$ components, and spacecraft subsystems in the networks of $1023$ and $2000$ components, with the binary \textbf{(middle)} and star \textbf{(right)} interconnection topologies, respectively. Green \protect\greensquare\ and pink \protect\pinksquare\ surfaces represent initial and unsafe level sets $\mathcal B (x) = \eta$ and $\mathcal B (x) = \mu$, respectively (due to the proximity of the level sets, the unsafe level set $\mathcal{B}(x)=\mu$, indicated in pink, is not clearly visible). Initial and unsafe regions are depicted by blue \protect\bluesquare\ and red \protect\redsquare\ boxes.} 
	\label{fig:2}
\end{figure*}

\subsection{Spacecraft network with star interconnection topology}
We consider the spacecraft network with the star topology, in which the spacecraft's dynamics for all $i \in \{2,\ldots,2000\}$  are characterized by
\begin{align}\notag
		\dot{x}_{i_1} &\!=\! \frac{J_{i_2} - J_{i_3}}{J_{i_1}}\, x_{i_2}\, x_{i_3} 
		\!\!+\!\! \frac{1}{J_{i_1}}\, u_{i_1}\!\! +\!\! \frac{0.02}{J_{i_1}}\, x_{1_1}, \\\notag 
		\dot{x}_{i_2} &\!=\! \frac{J_{i_3} - J_{i_1}}{J_{i_2}}\, x_{i_1}\, x_{i_3} 
		\!\! +\!\! \frac{1}{J_{i_2}}\, u_{i_2} \!\! +\!\! \frac{0.02}{J_{i_2}}\, x_{1_2}, \\\label{space-star} 
		\dot{x}_{i_3} &\!=\! \frac{J_{i_1} - J_{i_2}}{J_{i_3}}\, x_{i_1}\, x_{i_2} \!\! +\!\! \frac{1}{J_{i_3}}\, u_{i_3} \!\! +\!\! \frac{0.02}{J_{i_3}}\, x_{1_2},
\end{align}
while subsystem $i=1$ evolves as
\begin{align*}
		\dot{x}_{i_1} &= \frac{J_{i_2} - J_{i_3}}{J_{i_1}}\, x_{i_2}\, x_{i_3} 
		+ \frac{1}{J_{i_1}}\, u_{i_1}, \\
		\dot{x}_{i_2} &= \frac{J_{i_3} - J_{i_1}}{J_{i_2}}\, x_{i_1}\, x_{i_3} 
		+ \frac{1}{J_{i_2}}\, u_{i_2}, \\
		\dot{x}_{i_3} &= \frac{J_{i_1} - J_{i_2}}{J_{i_3}}\, x_{i_1}\, x_{i_2} 
		+ \frac{1}{J_{i_3}}\, u_{i_3}.
\end{align*}
Then, one can rewrite the dynamics in~\eqref{space-star} in the form of \eqref{sys2} with actual $\mathcal{R}_i(x_i) = [x_{i_1},x_{i_2},x_{i_3},   x_{i_2}x_{i_3}; x_{i_3}x_{i_1};\\ x_{i_1}x_{i_2} ]$ and
\begin{align*}
		A_i \!&=\! \begin{bmatrix}  0 \quad& 0 \quad& 0 \quad& \frac{J_{i_2} - J_{i_3}}{J_{i_1}} \quad& 0 \quad& 0  \\ 0 \quad& 0 \quad& 0 \quad& 0 \quad& \frac{J_{i_3} - J_{i_1}}{J_{i_2}} & 0 \\0 \quad& 0 \quad& 0 \quad& 0 \quad& 0 \quad& \frac{J_{i_1} - J_{i_2}}{J_{i_3}} \end{bmatrix}\!\!,\\
		B_i \!&= \! \begin{bmatrix} \frac{1}{J_{i_1}}  & 0  & 0\\ 0 &  \frac{1}{J_{i_2}} &0 \\ 0 &0 &\frac{1}{J_{i_3}}    \end{bmatrix}\!\!,
		~~~ D_i \!=\!\!  \begin{bmatrix} \frac{0.02}{J_{i_1}}  & 0  & 0\\ 0 &  \frac{0.02}{J_{i_2}} &0 \\ 0 &0 &\frac{0.02}{J_{i_3}}    \end{bmatrix}\!\!,\\
\end{align*}		
where both the matrices $A_i$ and $B_i$ are presumed to be unknown. While the precise structure of $\mathcal{R}_i(x_i)$ is also not known, the dictionary of monomials up to the degree of $2$ is given by \(\mathcal{R}_i(x_i)= [x_{i_1}; x_{i_2}; x_{i_3}; x_{i_2} x_{i_3}; x_{i_3} x_{i_1}; x_{i_1} x_{i_2};\\x^2_{i_1}; x^2_{i_2}; x^2_{i_3}]\). Under the \emph{star} topology as
\begin{align*}		
		\mathds{M}&=\{\mathbb{m}_{ij}\} = \begin{cases}\mathbf{0}_{3 \times 3},& i=j, \\ \mathbf{I}_3 , & j=1, i \neq j, \\ \mathbf{0}_{3 \times 3}, & \text {Otherwise},\end{cases}\\
\end{align*}
the matrices of the interconnected network can be constructed as 	
\begin{align*}		
	A(x)&=\left\{\mathbb{a}_{i j}\right\}= \begin{cases}A_i \Theta_i(x_i),& i=j, \\ D_{i}, & j=1, i \neq j, \\ \mathbf{0}_{3 \times 3}, & \text {Otherwise},\end{cases}\\
	B &= \mathsf{blkdiag}(B_1,\ldots,B_{2000}).
	\end{align*}
The regions of interest are given as the state set \(X_i = [-5, 5]^3\), the initial set \(X_{0_i} = [-2,2]^3\), and the unsafe set \(X_{a_i} = [2.5,5] \times [-5,-3] \times [-5,-4] \cup [2.5,5] \times [4,5] \times [2.5,5] \cup [-5,-4] \times [4, 5] \times [2.5,5]\) for all \(i \in \{1,\ldots,2000\}\).

\subsection{Chen network with line interconnection topology}
We analyze a Chen network with the line topology, in which each subsystem has $3$ state variables $x_i=\left[x_{i_1}; x_{i_2}; x_{i_3}\right],$ for all $i \in \{2,\ldots,1000\}$. The dynamics of each subsystem are described as
\begin{align}\notag
	& \dot{x}_{i_1}=35\, x_{i_2}-35\, x_{i_1} - 0.005 \, x_{{(i-1)}_1} + u_{i_1},\\\notag
	& \dot {x}_{i_2}=-7\, x_{i_1}+28\,x_{i_2}-x_{i_1} x_{i_3} -  0.005 \, x_{{(i-1)}_2}+ u_{i_2}, \\\label{chen-fully}
	& \dot{x}_{i_3}=x_{i_1} x_{i_2}-3\, x_{i_3} - 0.005\, x_{{(i-1)}_3}+ u_{i_3},
	\end{align}
while subsystem $i=1$ evolves as
\begin{align}\notag
		& \dot{x}_{i_1}=35\, x_{i_2}-35\, x_{i_1} + u_{i_1},\\\notag
		& \dot {x}_{i_2}=-7\, x_{i_1}+28\,x_{i_2}-x_{i_1} x_{i_3} + u_{i_2}, \\
		& \dot{x}_{i_3}=x_{i_1} x_{i_2}-3\, x_{i_3} + u_{i_3}.
\end{align}
One can rewrite the dynamics in~\eqref{chen-fully} in the form of \eqref{sys2} with actual $\mathcal{R}_i(x_i) = [x_{i_1}; x_{i_2}; x_{i_3}; x_{i_1}x_{i_3}; x_{i_1}x_{i_2}]$, and
	\begin{align*}
	A_i &= \begin{bmatrix} -35& 35 & 0 & 0 & 0  \\-7 & 28 & 0 & -1 & 0  \\ 0 & 0 & -3 & 0 & 1 \end{bmatrix}\!\!, \,
	B_i = \mathbf{I}_3,\, D_i = -0.005\, \mathbf{I}_{3}.\\
\end{align*}
The matrices \(A_i\) and \(B_i\), along with the exact form of \(\mathcal{R}_i(x_i)\), are unknown. The dictionary of monomials up to the degree of $2$ is given by \(\mathcal{R}_i(x_i)= [x_{i_1}; x_{i_2}; x_{i_3}; x_{i_1} x_{i_3}; x_{i_1}x_{i_2}; x_{i_2}x_{i_3}; x^2_{i_1}; x^2_{i_2}; x^2_{i_3}]\). Under the \emph{line} topology as
		\begin{align*}
    \mathds{M} &= \begin{bmatrix}
		\mathbf{0}_{3 \times 3} &&  \mathbf{0}_{3 \times 3}  &&  \mathbf{0}_{3 \times 3} &&  \cdots  &&  \mathbf{0}_{3 \times 3}\\
	    \mathbf{I}_{3} &&  \mathbf{0}_{3 \times 3}  && \mathbf{0}_{3 \times 3}  && \cdots  && \mathbf{0}_{3 \times 3} \\
	\vdots \!\! && \!\!\quad \!\! && \!\!\ddots \!\! && \!\!\quad \!\! && \!\!\vdots \\
		\mathbf{0}_{3 \times 3} && \cdots  && \mathbf{I}_{3}  &&   \mathbf{0}_{3 \times 3}  && \!\!\mathbf{0}_{3 \times 3} \\
		\mathbf{0}_{3 \times 3}  &&  \cdots  && \mathbf{0}_{3 \times 3}  &&   \mathbf{I}_{3}  &&  \mathbf{0}_{3 \times 3}
	\end{bmatrix}_{3000\times 3000}\!\!\!\!\!\!\!\!\!\!\!\!\!\!\!\!\!\!\!\!\!\!\!\!\!,
\end{align*}
the matrices of the interconnected network can be arranged as 	
\begin{align*}
		A(x)~\text{as in ~\eqref{A(x)-chen}}, ~~B= \mathsf{blkdiag}(B_1,\ldots,B_{1000}).
\end{align*}
\begin{figure*}[!t] 
		\rule{\textwidth}{0.1pt}
\begin{align}\label{A(x)-chen}
	A(x) &= \begin{bmatrix}
	A_1\Theta_1(x_1)  & \mathbf{0}_{3 \times 3} & \mathbf{0}_{3 \times 3}  & \cdots  &  \mathbf{0}_{3 \times 3}\\
	D_{2}  & A_2\Theta_2(x_2)  & \mathbf{0}_{3 \times 3}  &  \cdots  & \mathbf{0}_{3 \times 3} \\
	\vdots  & \!\! \!\! & \!\!\ddots \!\! & \!\! \!\! & \vdots \\
	\mathbf{0}_{3 \times 3}  & \cdots  & D_{999}   &  A_{999}\Theta_{999}(x_{999})  & \mathbf{0}_{3 \times 3} \\
	\mathbf{0}_{3 \times 3}  & \cdots  & \mathbf{0}_{3 \times 3}  & D_{1000}  & \!\!A_{1000} \Theta_{1000}(x_{1000})
\end{bmatrix}\!\!.
\end{align}
	\rule{\textwidth}{0.1pt}
\end{figure*}
The regions of interest are defined by the state set \(X_i = [-20, 20]^3\), the initial set \(X_{0_i} = [-2.5,2.5]^3\), and the unsafe set \(X_{a_i} = [-20,-9] \times[-20,-11]\times [-20,-8] \cup [3.5,20] \times[5,20] \times [4, 20]\) for all $i \in \{1,\ldots,1000\}$.

Closed-loop trajectories of representative Chen subsystems with line interconnection topology, and spacecraft subsystems with binary and star interconnection topologies are depicted in Figures \ref{fig:2}.

\begin{figure}
	\centering
	\begin{subfigure}[t]{0.24\textwidth}
		\includegraphics[width=\linewidth]{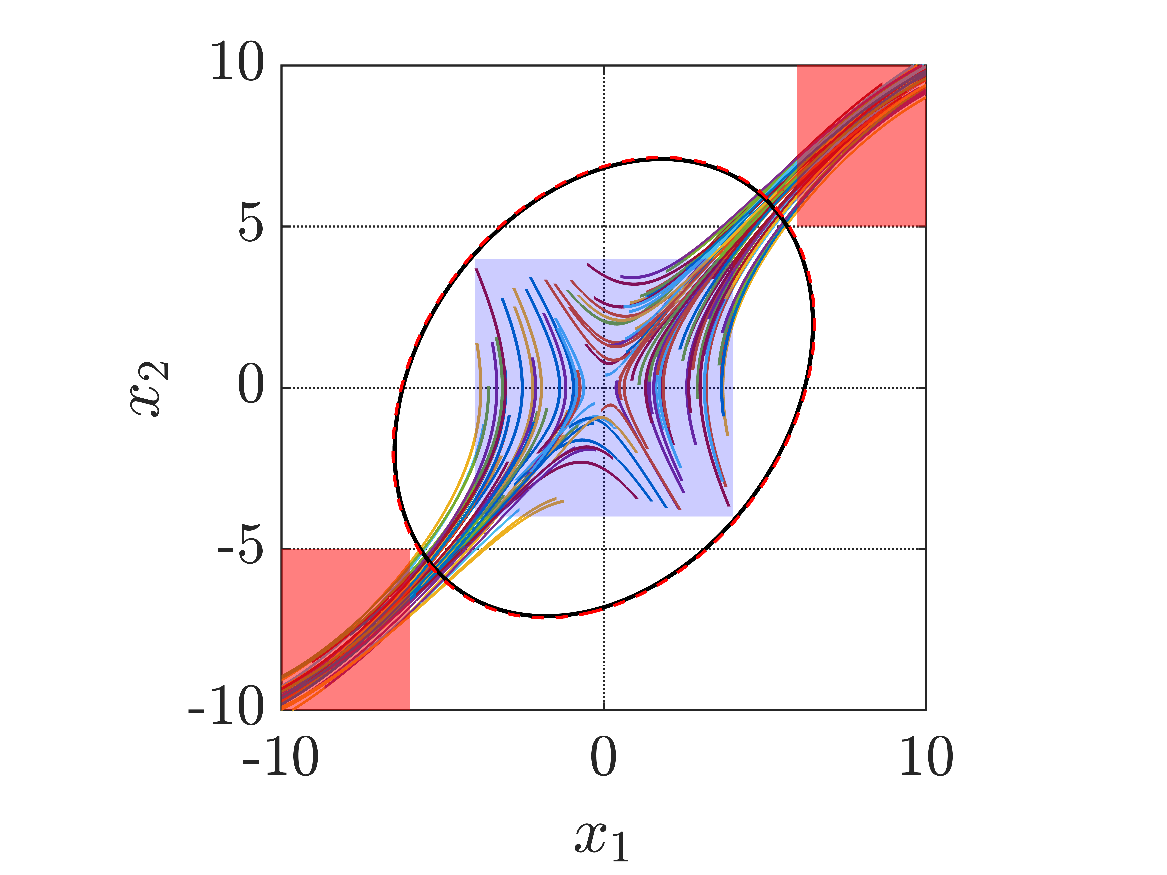}
		\caption{Open-loop trajectories}
	\end{subfigure}
	\hspace{-0.4cm}
	\begin{subfigure}[t]{0.24\textwidth}
		\includegraphics[width=\linewidth]{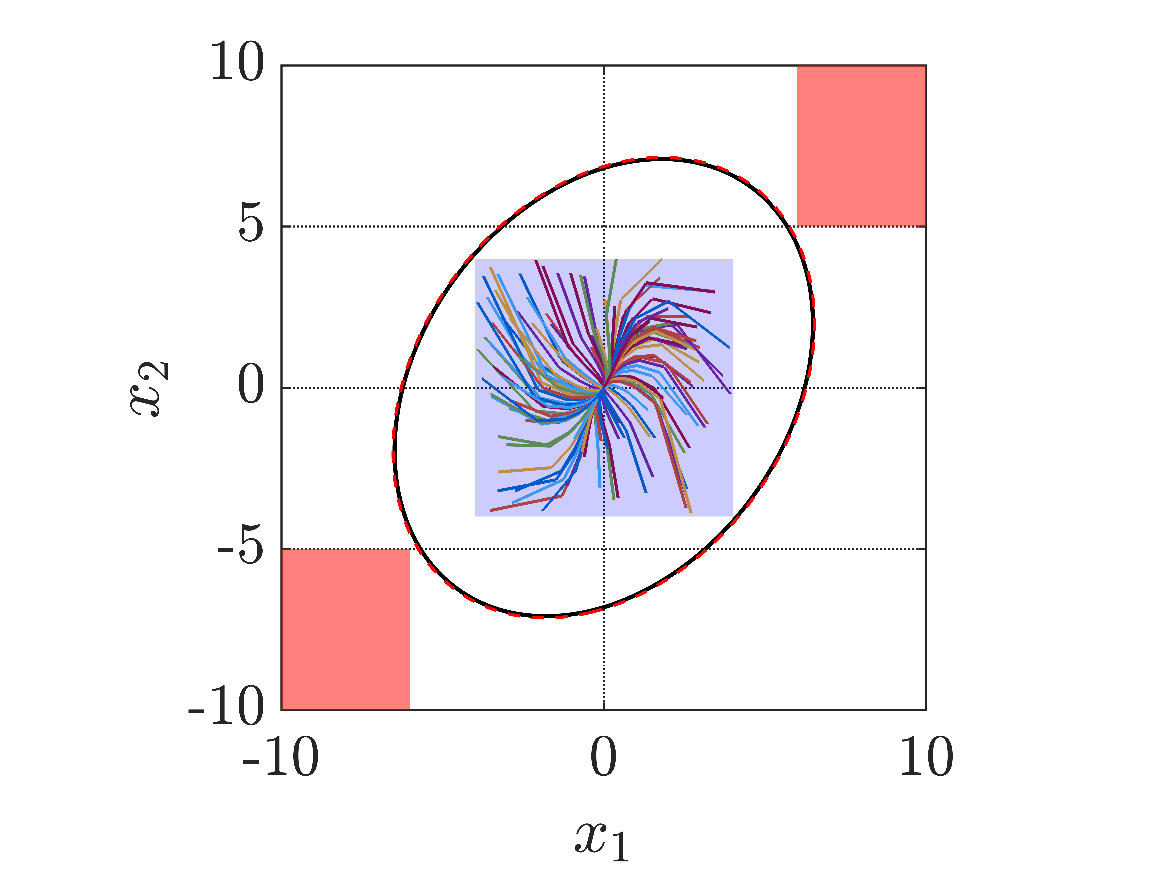}
		\caption{Closed-loop trajectories}
	\end{subfigure}
	\caption{Trajectories of $120$ representative
			Duffing oscillator subsystems in a network of $1000$ components with a binary interconnection topology. Moreover, $\mathcal{B}(x)=\eta$ and $\mathcal{B}(x)=\mu$ are indicated by~\sampleline{black, thick} and~\sampleline{dashed, red, thick}. Initial and unsafe regions are depicted by purple \protect\bluesquare\ and red \protect\redsquare\ boxes. As shown in (a), without a designed safety controller, the trajectories of representative subsystems immediately violate the safety specifications.}
	\label{fig:3}
\end{figure}

\subsection{Duffing oscillator network with binary  interconnection topology}
We analyze a Duffing oscillator network with the binary topology, where each subsystem has two state variables \(x_i=\left[x_{i_1}; x_{i_2}\right]\). The dynamics of each subsystem are given by
\begin{align}\notag
		\dot{x}_{i_1}&= x_{i_2} + u_{i_1},\\\notag
		\dot {x}_{i_2}\!&=\! 2\,x_{i_1}\!-0.5\, x_{i_2}\!-0.01\,x^3_{i_1} \\\label{duffing-binary} &\quad+ \!0.005\, \Omega\,(i=2 j \vee i=2 j+1) x_{j_1} + u_{i_2}, 
\end{align}
where $\vee$ stands for the logical $\mathsf{OR}$ operation and $\Omega$ is an indicator that we define it as
	\begin{equation*}
		\Omega\,(i=2 j \vee i=2 j+1)= \begin{cases}1, & i=2 j \vee i=2 j+1, \\ 0, & \text { Otherwise}.\end{cases}
\end{equation*}
One can rewrite the dynamics in~\eqref{duffing-binary} in the form of \eqref{sys2} with actual $\mathcal{R}_i(x_i) = [x_{i_1}; x_{i_2}; x_{i_1}^3]$ and
\begin{align*}
		A_i &= \begin{bmatrix} 0 & 1 & 0  \\ 2 & -0.5 & -0.01  \end{bmatrix}\!\!, \:
		B_i = \mathbf{I}_2,\:
		D_i = \begin{bmatrix} 0 & 0  \\ 0.005 &  0  \end{bmatrix}\!\!.
\end{align*}
The matrices \(A_i\) and \(B_i\), as well as the precise form of \(\mathcal{R}_i(x_i)\), are unknown. The dictionary of monomials up to the degree of $3$ is constructed as \(\mathcal{R}_i(x_i) = [x_{i_1}; x_{i_2}; x_{i_1}x_{i_2}; x^2_{i_1}; x^2_{i_2}; x^2_{i_1}x_{i_2}; x_{i_1}x^2_{i_2}; x^3_{i_1}; x^3_{i_2}]\). Under the binary topology as 
\begin{align*}
		\mathds{M}&=\{\mathbb{m}_{ij}\} = \begin{cases}  \mathbf{0}_{2 \times 2}, & i=j, \\ \mathbf{I}_2, & i=2 j \vee i=2 j+1, \\ \mathbf{0}_{2 \times 2}, & \text{Otherwise}, \end{cases}\\
\end{align*}
the matrices of the interconnected network can be constructed as
\begin{align*}
	A(x)&=\left\{\mathbb{a}_{i j}\right\} = \begin{cases} A_i\Theta_{i}(x_{i}), & i=j, \\ D_{i}, & i=2 j \vee i=2 j+1, \\ \mathbf{0}_{2 \times 2}, & \text{Otherwise}, \end{cases}\\
	B &= \mathsf{blkdiag}(B_1,\ldots,B_{1023}).
	\end{align*}
The regions of interest are defined as the state set \(X_i = [-10,10]^2\), the initial set \(X_{0_i} = [-4,4]^2\), and the unsafe set \(X_{a_i} = [-10,-6]\times[-10,-5]\cup [6,10]\times[5,10]\) for all \(i \in \{1,\ldots,1023\}\).

Trajectories of representative Duffing oscillator subsystems with the binary interconnection topology are depicted in Figure~\ref{fig:3}.

\section{Conclusion}\label{sec:Conclusion}
Our paper offered a compositional data-driven methodology with noisy data for designing \emph{fully-decentralized} safety controllers for large-scale networks with unknown mathematical models. By leveraging a single input-state trajectory for each unknown subsystem and applying dissipativity compositional reasoning, our method efficiently designed a control barrier certificate and its associated controller, ensuring the safety of the entire network through the construction of control storage certificates for individual subsystems from data. The key takeaway message is that our approach significantly reduces computational complexity from a polynomial to a linear scale in the number of subsystems, making it practical for ensuring safety certificates across large-scale networks with unknown models. Developing a compositional data-driven approach for safety certificates of large-scale networks that encompasses \emph{more general nonlinear systems} beyond polynomials is currently under investigation as future work.

\bibliographystyle{agsm}
\bibliography{biblio}	

\newpage

\appendix

\section{Appendix}
The thorough presentation of CSC $\mathcal{S}_i(x_i)$, local controllers \(u_i\), CBC $\mathcal{B}(x)$, safety controllers \(u\), and the network parameters, including \(\eta\), \(\mu\), and \(\lambda\), has been provided.
\subsection{Lorenz network with fully-interconnected topology}
\textbf{Subsystems: CSCs and local controllers.}
\begin{align*}
	u_{i_1}  & \!\!= 
	-4.3237\,x_{i_1}^2 + 12.7489\,x_{i_1}x_{i_2} - 3.3941\,x_{i_1}x_{i_3}\\
	\!\! & \!\!\quad - 3.5732\,x_{i_2}^2 + 1.9326\,x_{i_2}x_{i_3} + 4.4841\,x_{i_3}^2\\
	\!\! & \!\!\quad - 552.112\,x_{i_1} + 33.5012\,x_{i_2} + 138.1545\,x_{i_3} \\[1em]
	u_{i_2}  & \!\!= 
	-38.2124\,x_{i_1}^2 + 15.2553\,x_{i_1}x_{i_2} - 3.8784\,x_{i_1}x_{i_3}\\
	\!\! & \!\!\quad - 0.68332\,x_{i_2}^2 + 5.0421\,x_{i_2}x_{i_3} - 13.464\,x_{i_3}^2\\
	\!\! & \!\!\quad + 160.3144\,x_{i_1} - 165.9278\,x_{i_2} + 145.056\,x_{i_3} \\[1em]
	u_{i_3}  & \!\!= 
	9.5670\,x_{i_1}^2 + 2.8849\,x_{i_1}x_{i_2} - 19.0893\,x_{i_1}x_{i_3}\\
	\!\! & \!\!\quad - 6.0919\,x_{i_2}^2 + 13.9538\,x_{i_2}x_{i_3} + 2.4699\,x_{i_3}^2\\
	\!\! & \!\!\quad - 254.5260\,x_{i_1} - 140.9004\,x_{i_2} - 181.1822\,x_{i_3} \\[1em]
	\mathcal{S}_i(x_i)  & \!= 
	4.6884\,x_{i_1}^2 - 1.2131\,x_{i_1}x_{i_2} - 0.49461\,x_{i_1}x_{i_3}\\
	\!\! & \!\!\quad + 1.4175\,x_{i_2}^2 + 0.37467\,x_{i_2}x_{i_3} + 1.51\,x_{i_3}^2\\
	\Theta_i(x_i) &=   \begin{bmatrix}
		1 \quad& 0 \quad& 0 \quad& x_{i_3} & x_{i_2} & 0       & 0       & x_{i_1} & 0 \\
		0 \quad& 1 \quad& 0 \quad& 0       & 0       & x_{i_3} & x_{i_2} & 0       & 0 \\
		0 \quad& 0 \quad& 1 \quad& 0       & 0       & 0       & 0       & 0       & x_{i_3}
	\end{bmatrix}^\top
\end{align*}
\textbf{Network: CBC and safety controller.}
\begin{align*}
	\mathcal{B}(x) &=\sum_{i=1}^{1000} \mathcal{S}_i\left( x_i\right), ~ u = [u_1;\dots; u_{1000}]
	\\\eta &=\sum_{i=1}^{1000} \eta_i=9.82 \times 10^4,\, \mu=\sum_{i=1}^{1000} \mu_i=10^5\\  ~\lambda&=\underset{i}{\min}\{\lambda_i\}=0.99,~i \in\{1, \ldots, 1000\}
\end{align*}
\subsection{Lorenz network with ring interconnection topology}
\textbf{Subsystems: CSCs and local controllers.}
\begin{align*}
	u_{i_1} &= 
	-6.1695\,x_{i_1}^2 + 11.0212\,x_{i_1}x_{i_2} + 7.0979\,x_{i_1}x_{i_3}\\
	&\quad - 22.1475\,x_{i_2}^2 - 5.0455\,x_{i_2}x_{i_3} + 4.6617\,x_{i_3}^2\\
	&\quad - 1856.3628\,x_{i_1} + 30.1409\,x_{i_2} + 148.9849\,x_{i_3}
\end{align*}
\begin{align*}
	u_{i_2} &= 
	-14.8764\,x_{i_1}^2 + 126.4865\,x_{i_1}x_{i_2} + 33.1705\,x_{i_1}x_{i_3}\\
	&\quad - 4.4055\,x_{i_2}^2 - 6.5109\,x_{i_2}x_{i_3} - 41.6344\,x_{i_3}^2\\
	&\quad + 354.2923\,x_{i_1} - 333.6824\,x_{i_2} - 109.2299\,x_{i_3} \\[1em]
	u_{i_3} &= 
	-37.2251\,x_{i_1}^2 - 2.9349\,x_{i_1}x_{i_2} - 35.6788\,x_{i_1}x_{i_3}\\
	&\quad - 8.6932\,x_{i_2}^2 + 37.6969\,x_{i_2}x_{i_3} + 8.4038\,x_{i_3}^2\\
	&\quad - 80.8152\,x_{i_1} - 15.1855\,x_{i_2} - 319.0002\,x_{i_3} \\[1em]
	\mathcal{S}_i(x_i) &= 
	7.0168\,x_{i_1}^2 - 0.64833\,x_{i_1}x_{i_2} - 0.92392\,x_{i_1}x_{i_3}\\
	&\quad + 1.2164\,x_{i_2}^2 + 0.43125\,x_{i_2}x_{i_3} + 1.3117\,x_{i_3}^2\\
		\Theta_i(x_i) &=   \begin{bmatrix}
			1 \quad& 0 \quad& 0 \quad& x_{i_3} & x_{i_2} & 0       & 0       & x_{i_1} & 0 \\
			0 \quad& 1 \quad& 0 \quad& 0       & 0       & x_{i_3} & x_{i_2} & 0       & 0 \\
			0 \quad& 0 \quad& 1 \quad& 0       & 0       & 0       & 0       & 0       & x_{i_3}
		\end{bmatrix}^\top
\end{align*}
\textbf{Network: CBC and safety controller.}
\begin{align*}
	\mathcal{B}(x) & =\sum_{i=1}^{1500} \mathcal{S}_i\left( x_i\right), ~ u = [u_1;\dots; u_{1500}]
	\\\eta &=\sum_{i=1}^{1500} \eta_i=1.85 \times 10^5, ~\mu=\sum_{i=1}^{1500} \mu_i=1.89 \times 10^5\\
	~\lambda&=\underset{i}{\min}\{\lambda_i\}=0.99,~i \in\{1, \ldots, 1500\}
\end{align*}
\subsection{Spacecraft network with binary interconnection topology}
\textbf{Subsystems: CSCs and local controllers.}
\begin{align*}
	u_{i_1} & \!\!= 
	-69.7274\,x_{i_1}^2 + 10.0955\,x_{i_1}x_{i_2} + 234.4275\,x_{i_1}x_{i_3}\\
	\!\! & \!\!\quad +201.4344\,x_{i_2}^2 - 348.6678\,x_{i_2}x_{i_3} + 63.2356\,x_{i_3}^2\\
	\!\! & \!\!\quad -30845.1363\,x_{i_1} - 697.1356\,x_{i_2} + 1994.8684\,x_{i_3}\\[1em]
	u_{i_2}  & \!\!= 
	-44.35\,x_{i_1}^2 - 574.1542\,x_{i_1}x_{i_2} + 358.6686\,x_{i_1}x_{i_3}\\
	\!\! & \!\!\quad + 72.9450\,x_{i_2}^2 + 424.4316\,x_{i_2}x_{i_3} - 101.8604\,x_{i_3}^2\\
	\!\! & \!\!\quad + 636.2872\,x_{i_1} - 12422.2623\,x_{i_2} - 5416.1388\,x_{i_3}\\[1em]
	u_{i_3}  & \!\!= 
	-253.6507\,x_{i_1}^2 + 834.1136\,x_{i_1}x_{i_2} - 522.0015\,x_{i_1}x_{i_3}\\
	\!\! & \!\!\quad - 423.0909\,x_{i_2}^2 - 24.5980\,x_{i_2}x_{i_3} + 72.7423\,x_{i_3}^2\\
	\!\! & \!\!\quad + 1673.5721\,x_{i_1} - 4785.6901\,x_{i_2} - 23915.8383\,x_{i_3}\\[1em]
	\mathcal{S}_i(x_i)  & \!\!= 
	8.0256\,x_{i_1}^2 + 0.3716\,x_{i_1}x_{i_2} - 0.65001\,x_{i_1}x_{i_3}\\
	\!\! & \!\!\quad + 3.1004\,x_{i_2}^2 + 1.7535\,x_{i_2}x_{i_3} + 4.4171\,x_{i_3}^2\\
	\Theta_i(x_i) &= \begin{bmatrix}
		1 \quad& 0 \quad& 0 \quad& 0 & x_{i_3} & x_{i_2} & x_{i_1} & 0 & 0 \\
		0 \quad& 1 \quad& 0 \quad& 0 & 0       & 0       & 0   & x_{i_2} & 0 \\
		0 \quad& 0 \quad& 1 \quad& x_{i_2} & 0 & 0 & 0 & 0 & x_{i_3}
	\end{bmatrix}^\top
\end{align*}
\textbf{Network: CBC and safety controller.}
\begin{align*}
	\mathcal{B}(x) &=\sum_{i=1}^{1023} \mathcal{S}_i\left( x_i\right), ~ u = [u_1;\dots; u_{1023}]\\\eta &=\sum_{i=1}^{1023} \eta_i=8.92 \times 10^4,~ \mu=\sum_{i=1}^{1023} \mu_i=9.13 \times 10^4\\
	\lambda&=\underset{i}{\min}\{\lambda_i\}=0.99,~i \in\{1, \ldots, 1023\}
\end{align*}
\subsection{Spacecraft network with star interconnection topology}
\textbf{Subsystems: CSCs and local controllers.}
\begin{align*}
	u_{i_1} & \!\!= 
	-97.8321\,x_{i_1}^2 +  2.5578\,x_{i_1}x_{i_2} + 167.9864\,x_{i_1}x_{i_3}\\
	\!\! & \!\!\quad +100.5372\,x_{i_2}^2 -445.1097\,x_{i_2}x_{i_3} +126.8911\,x_{i_3}^2\\
	\!\! & \!\!\quad -36056.2582\,x_{i_1} -287.4528\,x_{i_2} +3598.8481\,x_{i_3}\\[1em]
	u_{i_2}  & \!\!= 
	101.8186\,x_{i_1}^2 -797.5359\,x_{i_1}x_{i_2} +322.5670\,x_{i_1}x_{i_3}\\
	\!\! & \!\!\quad +183.1781\,x_{i_2}^2 +670.2757\,x_{i_2}x_{i_3} -144.0050\,x_{i_3}^2\\
	\!\! & \!\!\quad + 161.9852\,x_{i_1} -13176.2469\,x_{i_2} -7986.3670\,x_{i_3}\\[1em]
	u_{i_3}  & \!\!= 
	-170.5333\,x_{i_1}^2 +1164.6939\,x_{i_1}x_{i_2} -546.8951\,x_{i_1}x_{i_3}\\
	\!\! & \!\!\quad -470.7482\,x_{i_2}^2 -281.0812\,x_{i_2}x_{i_3} + 91.0744\,x_{i_3}^2\\
	\!\! & \!\!\quad + 3141.7310\,x_{i_1} -11752.1641\,x_{i_2} -38649.4363\,x_{i_3}\\[1em]
	\mathcal{S}_i(x_i)  & \!\!= 
	8.8806\,x_{i_1}^2 + 0.12498\,x_{i_1}x_{i_2} - 1.6575\,x_{i_1}x_{i_3}\\
	\!\! & \!\!\quad + 3.3549\,x_{i_2}^2 + 3.7422\,x_{i_2}x_{i_3} + 6.4872\,x_{i_3}^2\\
		\Theta_i(x_i) &=\begin{bmatrix}
			1 \quad& 0 \quad& 0 \quad& 0 & x_{i_3} & x_{i_2} & x_{i_1} & 0 & 0 \\
			0 \quad& 1 \quad& 0 \quad& 0 & 0       & 0       & 0   & x_{i_2} & 0 \\
			0 \quad& 0 \quad& 1 \quad& x_{i_2} & 0 & 0 & 0 & 0 & x_{i_3}
		\end{bmatrix}^\top
\end{align*}
\textbf{Network: CBC and safety controller.}
\begin{align*}
	\mathcal{B}(x) &=\sum_{i=1}^{2000} \mathcal{S}_i\left( x_i\right), ~ u = [u_1;\dots; u_{2000}]\\
	\eta &= \sum_{i=1}^{2000} \eta_i=2.56 \times 10^5, ~\mu = \sum_{i=1}^{2000} \mu_i=2.65 \times 10^5\\
	\lambda&=\underset{i}{\min}\{\lambda_i\}=0.99,~i \in\{1, \ldots, 2000\}
\end{align*}
\subsection{Chen network with line interconnection topology}
\textbf{Subsystems: CSCs and local controllers.}
\begin{align*}
	u_{i_1} &= 
	-4.525\,x_{i_1}^2 - 12.4249\,x_{i_1}x_{i_2} - 26.6267\,x_{i_1}x_{i_3}\\
	&\quad - 27.0846\,x_{i_2}^2 - 10.4337\,x_{i_2}x_{i_3} + 30.3557\,x_{i_3}^2\\  &\quad - 2252.0043\,x_{i_1} - 640.906\,x_{i_2} + 613.5597\,x_{i_3}
\end{align*}
\begin{align*}	
	u_{i_2} &= 
	-6.0461\,x_{i_1}^2 + 117.7663\,x_{i_1}x_{i_2} + 117.0708\,x_{i_1}x_{i_3}\\
	&\quad + 21.6111\,x_{i_2}^2 - 6.2939\,x_{i_2}x_{i_3} - 90.1604\,x_{i_3}^2\\
	&\quad + 1240.9254\,x_{i_1} - 342.1339\,x_{i_2} - 383.8634\,x_{i_3}\\[0.75em]
	u_{i_3} &= 
	-20.2585\,x_{i_1}^2 - 5.4401\,x_{i_1}x_{i_2} - 85.8727\,x_{i_1}x_{i_3}\\
	&\quad + 34.5787\,x_{i_2}^2 + 94.8588\,x_{i_2}x_{i_3} - 8.23\,x_{i_3}^2\\
	&\quad - 4.0755\,x_{i_1} + 577.6729\,x_{i_2} - 429.9882\,x_{i_3}\\[0.75em]
	\mathcal{S}_i(x_i) &= 
	2.3766\,x_{i_1}^2 + 0.52522\,x_{i_1}x_{i_2} - 0.92617\,x_{i_1}x_{i_3}\\
	&\quad + 0.67313\,x_{i_2}^2 - 0.3976\,x_{i_2}x_{i_3} + 0.47854\,x_{i_3}^2\\
		\Theta_i(x_i) &= \begin{bmatrix}
			1 \quad& 0 \quad& 0 \quad& x_{i_3} & x_{i_2} & 0 & 0 & x_{i_1} & 0 \\
			0 \quad& 1 \quad& 0 \quad& 0       & 0       & x_{i_3} & x_{i_2} & 0 & 0 \\
			0 \quad& 0 \quad& 1 \quad& 0       & 0       & 0 & 0 & 0 & x_{i_3}
		\end{bmatrix}^\top
\end{align*}
\textbf{Network: CBC and safety controller.}\vspace{-0.2cm}
\begin{align*}
	\mathcal{B}(x) &=\sum_{i=1}^{1000} \mathcal{S}_i\left( x_i\right), ~ u = [u_1;\dots; u_{1000}]
	\\\eta &=\sum_{i=1}^{1000} \eta_i=3.56 \times 10^4,\, \mu=\sum_{i=1}^{1000} \mu_i=3.62 \times 10^4\\
	~\lambda&=\underset{i}{\min}\{\lambda_i\}=0.99,~i \in\{1, \ldots, 1000\}
\end{align*}
\subsection{Duffing oscillator network with binary  interconnection topology}\vspace{-0.2cm}
\begin{align*}
	u_{i_1} &= 
	-1.4851\,x_{i_1}^3 - 11.1237\,x_{i_1}^2x_{i_2} - 6.8794\,x_{i_1}x_{i_2}^2\\
	&\quad - 6.3278\,x_{i_2}^3 + 3.7858\,x_{i_1}^2 - 15.4314\,x_{i_1}x_{i_2}\\
	&\quad - 2.3591\,x_{i_2}^2 - 417.6319\,x_{i_1} + 90.3340\,x_{i_2}\\[1em]
	u_{i_2} &= 
	19.4387\,x_{i_1}^3 - 8.1903\,x_{i_1}^2x_{i_2} + 12.0891\,x_{i_1}x_{i_2}^2\\
	&\quad - 7.3353\,x_{i_2}^3 + 24.1219\,x_{i_1}^2 - 6.9668\,x_{i_1}x_{i_2}\\
	&\quad + 0.6629\,x_{i_2}^2 + 124.2233\,x_{i_1} - 356.4827\,x_{i_2}\\[1em]
	\mathcal{S}_i(x_i) &= 
	10.4512\,x_{i_1}^2 - 5.3106\,x_{i_1}x_{i_2} + 8.7529\,x_{i_2}^2\\
\Theta_i(x_i) &=	\begin{bmatrix}
	1 \quad& 0 \quad& x_{i_2} \quad& x_{i_1} & 0 & 0 & x_{i_2}^{2} & x_{i_1}^{2} & 0 \\
	0 \quad& 1 \quad& 0       \quad& 0       & x_{i_2} & x_{i_1}^{2} & 0 & 0 & x_{i_2}^{2}
\end{bmatrix}^\top
\end{align*}
\textbf{Network: CBC and safety controller.}\vspace{-0.2cm}
\begin{align*}
	\mathcal{B}(x) &=\sum_{i=1}^{1023} \mathcal{S}_i\left( x_i\right), ~ u = [u_1;\dots; u_{1023}]	\\\eta &=\sum_{i=1}^{1023} \eta_i=4.16 \times 10^5,\, \mu=\sum_{i=1}^{1023} \mu_i=4.22 \times 10^5\\
	~\lambda&=\underset{i}{\min}\{\lambda_i\}=0.99,~i \in\{1, \ldots, 1023\}
\end{align*}

\end{document}